\documentclass[11pt]{article}

\usepackage{amsmath,amssymb,amsthm}
\usepackage{graphicx,cite,booktabs,microtype}
\usepackage{caption,placeins,xcolor}
\usepackage{titlesec}
\usepackage[titletoc,title]{appendix}
\usepackage{subcaption}
\usepackage[hidelinks]{hyperref}

\graphicspath{{figures/}}


\newtheorem{proposition}{Proposition}

\newcommand{\hoch}[1]{$\, ^{#1}$}
\newcommand{\Mpl}{M_{\rm Pl}}
\newcommand{\dd}{\mathrm d}

\begin{document}

	\begin{center}

{\Large\bfseries\boldmath Jordan-Frame Dynamics of Curvature Inflation: A Derivative Expansion of the Friedmann Constraint}

		\vspace{20pt}

		H.~Khodabakhshi\hoch{1}, F.~Shojai\hoch{2}

		\vspace{20pt}

		{\it \hoch{1}Center for Joint Quantum Studies and Department of Physics,\\
			School of Science, Tianjin University, Tianjin 300350, China}
		\medskip

		{\it \hoch{2}Department of Physics, University of Tehran,\\
			P.O. Box 14395-547, Tehran, Iran}

		\vspace{20pt}

\begin{abstract}
We derive a slow-roll approximation for the Jordan-frame Hubble rate $H_J$ in metric $f(R)$ gravity. A derivative expansion of the vacuum Friedmann constraint to third order gives $\dot H_J$ as a function of $H_J$. In the Starobinsky model we also show that the difference in $\dot H_J$ between two expanding solutions decreases as they pass through successively lower values of $H_J$, and that a signed square root of $H_J$ evolves smoothly through the zeros that follow inflation. We test the approximation in a regularized inverse-logarithmic $R^2$ model and in $R+\alpha^{p-1}R^p$, taking the pivot from the exact background 60 Einstein-frame e-folds before the end of inflation. For the inverse-logarithmic coupling $\delta=3$, the third-order approximation reproduces $\dot H_J$ at the pivot with a relative error of $3\times10^{-8}$. Integrating the approximate branch between the pivot and exit values of $H_J$ taken from the exact background gives 59.98 e-folds, compared with 60 on the exact trajectory. At the six couplings $\delta=0,1,\ldots,5$, the absolute difference from the numerical mode result is below $1.1\times10^{-5}$ in the scalar tilt $n_s$, and the relative difference is below $0.0052\%$ in the tensor-to-scalar ratio $r$. For these six benchmarks, the second-order spectral formulas account for nearly all of the difference from the numerical mode results. For $p=1.8$ the approximation itself adds to the error, which becomes $3.9\times10^{-5}$ in $n_s$ and $0.047\%$ in $r$. For the inverse-logarithmic benchmark with $\delta=3$, perturbative scalaron decay into a minimally coupled Higgs doublet places the pivot 51.59 e-folds before the end of inflation and gives $n_s=0.9690$ and $r=0.0067$. At fixed $\alpha\mu_0^2=0.006065$ and with the same decay rate prescription, $\delta\simeq8.7$ gives the central $n_s$ value inferred from ACT DR6, Planck, CMB lensing and DESI baryon acoustic oscillations. This comparison uses the published one-dimensional constraint.
\end{abstract}

	\end{center}
\vfill
\noindent{\small h\_khodabakhshi@tju.edu.cn \qquad fshojai@ut.ac.ir}
\thispagestyle{empty}
\pagebreak

\tableofcontents
\addtocontents{toc}{\protect\setcounter{tocdepth}{2}}

\section{Introduction}
\label{sec:intro}

In the Starobinsky model, the $R^2$ term adds a scalar degree of freedom to general relativity~\cite{Starobinsky1980,Vilenkin1985}. In the Einstein frame this field, the scalaron, rolls down a potential that is nearly flat at large field values and quadratic near its minimum~\cite{Whitt1984,Maeda1989}. This description makes it straightforward to calculate the inflationary spectra~\cite{Sotiriou2010,DeFelice2010}. In the Jordan frame the same degree of freedom appears in the time dependence of the curvature. We ask how much of the inflationary evolution and its perturbation spectrum can be inferred directly from the Hubble rate in that frame. We restrict the analysis to metric $f(R)$ gravity. Other quadratic curvature invariants can introduce additional degrees of freedom~\cite{Stelle1977,Stelle1978}.

The quadratic model provides a useful starting point because its vacuum Friedmann constraint is an exact second-order equation for $H_J$. During slow roll, $\dot H_J\simeq-M_\phi^2/6$, where $M_\phi^2=1/(6\alpha)$. Inflation is a transient attractor~\cite{Maeda1988}. After inflation, $H_J$ oscillates and its mean falls as $2/(3t_J)$~\cite{Mijic1986,DeFelice2010}. We compare solutions when they reach the same $H_J$. This removes the freedom to shift the origin of time and gives an exact expression for the decay of the difference in their velocities. The same Hubble equation appears singular at $H_J=0$. Writing $H_J=Y^2$ gives a regular equation for a signed variable $Y$ and explains how an expanding solution passes through each zero.

For a general $f(R)$, the coefficient of $\ddot H_J$ depends on $\dot H_J$ through $R=12H_J^2+6\dot H_J$. Expanding about $R_0=12H_J^2$ gives a slow-roll velocity that can be evaluated at any specified $H_J$ on the branch. We test this expansion in two settings. In a regularized inverse-logarithmic $R^2$ model the higher curvature derivatives are small during inflation. In a power-law deformation, $R+\alpha^{p-1}R^p$ with $p<2$, their effect is larger. For each model we compare the approximate trajectory with the exact background. We also use the frame transformation to calculate $n_s$ and $r$ on the slow-roll branch and compare them with a direct integration of the perturbations at the same pivot.

The inverse-logarithmic model also allows us to study the sensitivity of the spectra to the shape of the plateau. Logarithmic corrections to quadratic inflation have been studied in refs.~\cite{BenDayan2014,Rinaldi2015}. At large curvature the quadratic part of our action has the inverse-logarithmic dependence found in running quadratic gravity~\cite{LiuQuintinAfshordi2026}. We retain the Einstein--Hilbert term and choose a continuation that is smooth at $R=0$. These choices affect the potential slope during inflation and the scalaron mass at its minimum. Both enter the prediction at a specified comoving scale.

The observational motivation comes from the scalar tilt. The combination of ACT DR6, Planck, CMB lensing and DESI baryon acoustic oscillations (P-ACT-LB) gives $n_s=0.9743\pm0.0034$~\cite{ACTDR6_2025,ACTDR6ext_2025}, compared with $0.9649\pm0.0042$ from Planck 2018~\cite{Planck2018}. The implications for quadratic inflation depend on reheating and the cosmological parameters allowed in the fit~\cite{BezerraSobrinho2025}. We give results both at fixed e-fold number and for perturbative scalaron decay into the Higgs field. Reheating can involve nonperturbative particle production~\cite{Kofman1994,Kofman1997}. Our calculation assumes sudden thermalization after perturbative decay. We compare our predictions with published one-dimensional constraints.

Some of the results for the quadratic model are known. The slow-roll drift and the late-time mean of $H_J$ are given in refs.~\cite{Mijic1986,DeFelice2010}, the attractor behaviour is discussed in ref.~\cite{Maeda1988}, and the leading relation $r\simeq48\epsilon_J^2$ is in ref.~\cite{DeFelice2010}. What is new here is the exact identity for two solutions compared at equal $H_J$ (section~\ref{sec:contraction}) and the signed variable $Y$, which carries a solution through each zero of $H_J$ (section~\ref{sec:signed_root}). For general $f(R)$ we give the third-order expansion of the Friedmann constraint and solve it on the slow-roll branch (section~\ref{sec:general}). We then test this branch against exact backgrounds and numerical mode functions in two models (sections~\ref{sec:model}--\ref{sec:perturbations}), and for the inverse-logarithmic model we fix the pivot by a self-consistent reheating calculation. Appendix~\ref{app:convergence} gives the perturbation equations and numerical checks. We use the metric signature $(-,+,+,+)$.

\section{The quadratic model in the Hubble variable}
\label{sec:quadratic}

\subsection{Field equations}

The metric action is
\begin{equation}
 S=\frac{\Mpl^2}{2}\int \dd^4x\sqrt{-g_J}\,f(R),
 \qquad \Mpl^{-2}=8\pi G,
 \label{eq:action}
\end{equation}
on a spatially flat FLRW background. We write $R=R_J$ and $F=f_R$, and a dot means $\dd/\dd t_J$ unless stated otherwise. The vacuum equations include
\begin{align}
 3FH_J^2&=U-3H_J\dot F,
 &U&=\frac{RF-f}{2},\label{eq:constraint}\\
 R&=6(\dot H_J+2H_J^2),
 &\ddot F+3H_J\dot F+\frac{2f-RF}{3}&=0.
 \label{eq:ricci_trace}
\end{align}
The Ricci and trace equations propagate the constraint. For $\mathcal Z\equiv3FH_J^2-U+3H_J\dot F$ they give
\begin{equation}
 \dot{\mathcal Z}=-4H_J\mathcal Z.
 \label{eq:propagation}
\end{equation}
Away from $H_J=0$, differentiating the constraint gives back the spatial equation. At a zero of $H_J$ one should keep the trace and Ricci equations instead of dividing the constraint by $H_J$.

For $f=R+\alpha R^2$ with $\alpha>0$ we have $F=1+2\alpha R$, $U=\alpha R^2/2$ and $\dot F=2\alpha\dot R$. With $R=6(\dot H_J+2H_J^2)$ and $\dot R=6(\ddot H_J+4H_J\dot H_J)$, the constraint becomes, without approximation,
\begin{equation}
 36\alpha H_J\ddot H_J+108\alpha H_J^2\dot H_J
 -18\alpha\dot H_J^2+3H_J^2=0.
 \label{eq:quadratic_H}
\end{equation}
We write it as
\begin{equation}
 m\ddot H_J+\gamma\dot H_J-\beta\dot H_J^2+\mathcal G=0,
 \qquad
 m=36\alpha H_J,\quad \gamma=108\alpha H_J^2,\quad
 \beta=18\alpha,\quad \mathcal G=3H_J^2.
 \label{eq:mechanical}
\end{equation}
On a decreasing branch, the linear term contributes $-\gamma\dot H_J>0$ to $m\ddot H_J$, while the quadratic term contributes $+\beta\dot H_J^2$ for either sign of the velocity. We use mechanical terminology to distinguish these terms~\cite{Khodabakhshi2024}. Their coefficients depend on the normalization of the equation.

Define the Jordan-frame Hubble-flow parameters
\begin{equation}
 \epsilon_J=-\frac{\dot H_J}{H_J^2},\qquad
 \epsilon_{2J}=\frac{\dot\epsilon_J}{H_J\epsilon_J}
 =\frac{\ddot H_J}{H_J\dot H_J}+2\epsilon_J.
 \label{eq:flow}
\end{equation}
On an interval with $H_J>0$ and $\dot H_J<0$ the term ratios are, exactly,
\begin{equation}
 \mathcal I\equiv\frac{|m\ddot H_J|}{|\gamma\dot H_J|}
 =\frac{|\epsilon_{2J}-2\epsilon_J|}{3},\qquad
 \frac{\beta\dot H_J^2}{|\gamma\dot H_J|}=\frac{\epsilon_J}{6}.
 \label{eq:ratios}
\end{equation}
Both are small on the relaxed slow-roll branch. Dropping them gives the familiar drift (eqs.~(3.9) and (3.12) of ref.~\cite{DeFelice2010})
\begin{equation}
 \dot H_J\simeq-\frac{1}{36\alpha},\qquad
 \epsilon_J\simeq\frac{1}{36\alpha H_J^2}.
 \label{eq:drift}
\end{equation}
These ratios measure the size of the omitted terms along one solution. To determine whether nearby solutions approach the slow-roll branch, we must also compare their velocities.

\subsection{Comparison at equal Hubble rate}
\label{sec:contraction}

On a monotonic interval let $v(H_J)=\dot H_J$. Equation~\eqref{eq:quadratic_H} becomes
\begin{equation}
 \frac{\dd v}{\dd H_J}=-3H_J+\frac{v}{2H_J}-\frac{H_J}{12\alpha v}.
 \label{eq:phase_velocity}
\end{equation}
If two solutions $v_1,v_2<0$ are compared at the same $H_J>0$, their difference obeys the exact identity
\begin{equation}
 \frac{\dd(v_1-v_2)}{\dd H_J}
 =\left(\frac{1}{2H_J}+\frac{H_J}{12\alpha v_1v_2}\right)(v_1-v_2).
 \label{eq:finite_contraction}
\end{equation}
The coefficient is positive, so $|v_1-v_2|$ decreases as $H_J$ decreases. Comparing at equal $H_J$ removes the arbitrary time offset between two histories. For an infinitesimal displacement, with $\mathcal N_J=\ln a_J$,
\begin{equation}
 \frac{\dd\ln|\delta v|}{\dd\mathcal N_J}
 =\frac{v}{2H_J^2}+\frac{1}{12\alpha v}<0.
 \label{eq:local_contraction}
\end{equation}
This contraction holds on any common interval where both solutions expand and $H_J$ decreases. It gives a local account of attraction on that branch. Along the slow-roll drift, the rate approaches $-3$, consistent with the transient attractor of ref.~\cite{Maeda1988}.

To resolve the subleading drift, put $h=\sqrt\alpha H_J$ and $v=-(36\alpha)^{-1}w(h)$. Equation~\eqref{eq:phase_velocity} gives
\begin{equation}
 2h w\frac{\dd w}{\dd h}-w^2+216h^2(1-w)=0,
 \label{eq:w_equation}
\end{equation}
whose branch with $w\to1$ at large $h$ is
\begin{equation}
 \dot H_J=-\frac{1}{36\alpha}
 \left[1-\frac{1}{216\alpha H_J^2}
 +\frac{1}{7776\alpha^2H_J^4}
 +O\!\left((\alpha H_J^2)^{-3}\right)\right].
 \label{eq:drift_series}
\end{equation}
Consequently
\begin{equation}
 \mathcal I=\frac{\epsilon_J^2}{9}+O(\epsilon_J^3),
 \label{eq:inertia_series}
\end{equation}
so on this branch the inertial term is second order in $\epsilon_J$.

For a numerical check we integrate the exact Starobinsky equations from $x=\alpha R=100$ with the initial drift specified in section~\ref{sec:numerical}. The solution evolves for 87.5 Einstein-frame e-folds before reaching our reference event, 60 e-folds before the first crossing of $\epsilon_E=1$. At this event we find
\begin{equation}
 h_*=1.861059,\qquad \alpha\dot H_{J,*}=-0.02774094,\qquad
 \mathcal I_*=7.03\times10^{-6}.
 \label{eq:star_reference}
\end{equation}
Equation~\eqref{eq:drift_series} reproduces $\dot H_{J,*}$ with a relative error of $1.5\times10^{-7}$. The measured $\mathcal I_*$ is $1.3\%$ below $\epsilon_J^2/9=7.13\times10^{-6}$. Jordan-frame acceleration ends about 0.18 Einstein-frame e-fold after Einstein-frame acceleration. At the last sample before the Jordan exit we find $h_J=0.157919$ and $\mathcal I=0.0497$. At the exit itself the quadratic term is exactly one sixth of the linear term, by eq.~\eqref{eq:ratios}.

\begin{figure}[!htbp]
\centering
\includegraphics[width=\linewidth]{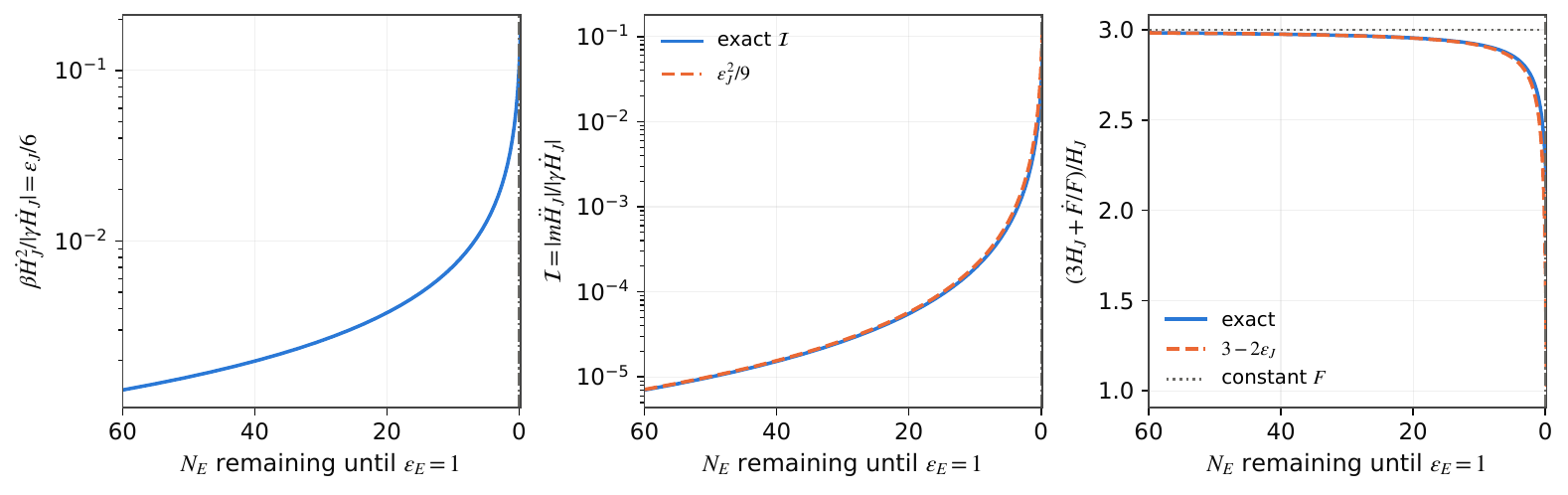}
\caption{Exact Starobinsky trajectory, plotted against the number of Einstein-frame e-folds left until the first $\epsilon_E=1$ crossing (dash-dotted line). Left: quadratic-to-linear ratio, equal to $\epsilon_J/6$. Middle: inertial-to-linear ratio $\mathcal I$ and its slow-roll estimate $\epsilon_J^2/9$. Right: tensor damping coefficient of eq.~\eqref{eq:tensor_eom} divided by $H_J$, its high-curvature estimate $3-2\epsilon_J$, and the constant-$F$ value 3. The curves end at the Jordan-frame exit, $\epsilon_J=1$.}
\label{fig:star_ratios}
\end{figure}

\subsection{Signed square root and Hubble zeroes}
\label{sec:signed_root}

On $H_J>0$ the substitution $H_J=Y^2$ removes the quadratic velocity term. Equation~\eqref{eq:quadratic_H} becomes
\begin{equation}
 24\alpha\ddot Y+72\alpha Y^2\dot Y+Y=0.
 \label{eq:signed_root}
\end{equation}
The positive root $Y=\sqrt{H_J}$ is only a local coordinate, but eq.~\eqref{eq:signed_root} is smooth at $Y=0$ and can be used with a signed $Y$. Its two signs describe the same $H_J$ away from a zero, so this is a regularizing parametrization, not a globally invertible field redefinition. The exact balance
\begin{equation}
 E_Y=12\alpha\dot Y^2+\frac12Y^2,
 \qquad \dot E_Y=-72\alpha Y^2\dot Y^2\leq0
 \label{eq:Y_energy}
\end{equation}
bounds $Y$ and $\dot Y$ in forward time for finite initial data.

To see what a zero of $Y$ means, use the regular trace--Ricci system
\begin{equation}
 \dot H_J=\frac R6-2H_J^2,\qquad
 \dot R=P,\qquad
 \dot P=-3H_JP-\frac{R}{6\alpha}.
 \label{eq:regular_star}
\end{equation}
At $H_J=0$ the constraint requires $R=\dot H_J=0$ but allows $P_0\neq0$. For an expanding tangency
\begin{equation}
 H_J(t_0+\Delta t_J)=\frac{P_0}{12}\Delta t_J^2-\frac{P_0}{864\alpha}\Delta t_J^4-\frac{P_0^2}{288}\Delta t_J^5+O(\Delta t_J^6),
 \qquad P_0>0.
 \label{eq:tangency}
\end{equation}
The trace equation gives $\dot P(t_0)=0$ and $\ddot P(t_0)=-P_0/(6\alpha)$. The damping term in eq.~\eqref{eq:signed_root} first contributes to $Y$ at order $\Delta t_J^4$, and hence to $H_J$ at order $\Delta t_J^5$. Since $P_0=12\dot Y_0^2$, the signed solution crosses zero with finite slope. Taking the positive root on both sides would instead give a cusp proportional to $|\Delta t_J|$. For $H_J=R=P=0$, uniqueness in eq.~\eqref{eq:regular_star} gives Minkowski space.

At small amplitude the signed variable oscillates with frequency $\omega_Y=1/\sqrt{24\alpha}$. Since
\begin{equation}
 R=12Y\dot Y+12Y^4,
 \label{eq:R_signed_root}
\end{equation}
a harmonic $Y\simeq A\cos(\omega_Yt_J+\theta)$ gives a curvature oscillation with frequency $2\omega_Y=1/\sqrt{6\alpha}$, the scalaron mass. Averaging eq.~\eqref{eq:Y_energy} over a period gives $\dot A\simeq-3A^3/8$ and, at late times,
\begin{equation}
 H_J\simeq\frac{4}{3(t_J-t_c)}
 \cos^2\!\left(\frac{t_J}{2\sqrt{6\alpha}}+\theta\right),
 \qquad \langle H_J\rangle\simeq\frac{2}{3(t_J-t_c)}.
 \label{eq:late_H}
\end{equation}
Averaging the energy balance thus recovers the late-time vacuum solution of ref.~\cite{Mijic1986}, also given in eq.~(3.31) of ref.~\cite{DeFelice2010}.

\section{\texorpdfstring{General $f(R)$: frames, derivative expansion and the slow-roll branch}{General f(R): frames, derivative expansion and the slow-roll branch}}
\label{sec:general}

\subsection{Frame transformation}

On a branch with $F>0$ and locally invertible $F(R)$, the exact transformation is
\begin{equation}
 g^E_{\mu\nu}=F g^J_{\mu\nu},\quad
 \phi=\sqrt{\frac32}\Mpl\ln F,\quad
 \dd t_E=\sqrt F\,\dd t_J,\quad
 H_E=\frac{H_J+\dot F/(2F)}{\sqrt F}.
 \label{eq:frames}
\end{equation}
The potential and the homogeneous equations are~\cite{Whitt1984,Maeda1989,DeFelice2010}
\begin{align}
 V_E&=\frac{\Mpl^2}{2}\frac{RF-f}{F^2},
 &V_{E,\phi}&=\frac{\Mpl}{\sqrt6}\frac{2f-RF}{F^2},\label{eq:potential_general}\\
 \frac{\dd^2\phi}{\dd t_E^2}+3H_E\frac{\dd\phi}{\dd t_E}+V_{E,\phi}&=0,
 &3\Mpl^2H_E^2&=\frac12\left(\frac{\dd\phi}{\dd t_E}\right)^2+V_E.
 \label{eq:EF_background}
\end{align}
The scalaron equation follows from the trace equation, and the Einstein-frame Friedmann equation from the transformed constraint~\eqref{eq:constraint}. For the quadratic model
\begin{equation}
 V_E=\frac{\Mpl^2}{8\alpha}
 \left(1-e^{-\sqrt{2/3}\,\phi/\Mpl}\right)^2,
 \qquad M_\phi^2=\frac{1}{6\alpha}.
 \label{eq:star_potential}
\end{equation}

Equation~\eqref{eq:frames} gives $H_J=\sqrt F\,[H_E-(\dd\phi/\dd t_E)/(\sqrt6\Mpl)]$. If $V_E\geq0$, the Einstein-frame constraint gives $H_E\geq|\dd\phi/\dd t_E|/(\sqrt6\Mpl)$ on an expanding branch, so $H_J\geq0$. This holds for any $f(R)$ with $F>0$ and $V_E\geq0$, including the model of section~\ref{sec:model}. For a nontrivial expanding solution, $H_J=0$ requires $V_E=0$ with $\dd\phi/\dd t_E>0$. For the quadratic model this is the tangency of section~\ref{sec:signed_root}, seen from the Einstein frame.

The scalaron mass follows from the curvature of its potential. We restrict the calculation to $F>0$, for a positive tensor kinetic term, and $f_{RR}>0$, for a locally invertible curvature branch~\cite{DolgovKawasaki2003}. Linearizing the trace equation about a constant-curvature vacuum $R_\star$ satisfying $2f=R_\star F$ gives the Jordan-frame mass
\begin{equation}
 M_{R,J}^2=\frac{F(R_\star)-R_\star f_{RR}(R_\star)}{3f_{RR}(R_\star)},
 \label{eq:stationary_mass}
\end{equation}
and the canonical Einstein-frame mass is $M_{R,J}^2/F(R_\star)$.

\subsection{Derivative expansion}
\label{sec:expansion}

For a general smooth $f(R)$ the exact Hubble equation is
\begin{equation}
 \mathcal E_{\rm ex}=18H_J f_{RR}(R)\ddot H_J
 +72H_J^2 f_{RR}(R)\dot H_J+3H_J^2f_R(R)-U(R)=0,
 \label{eq:general_exact}
\end{equation}
where $R=R_0+6\dot H_J$ and $R_0=12H_J^2$. Solving this constraint for $\ddot H_J$ gives the phase-space equation used in ref.~\cite{LiuQuintinAfshordi2026}. The dependence on the velocity enters both explicitly and through the curvature argument of $f$. We expand the latter dependence about $R_0$.

On an interval with $H_J\neq0$ write $v=\dot H_J$ and $a=\ddot H_J$, and count
\begin{equation}
 \frac{v}{H_J^2}=O(\epsilon_{\rm d}),\qquad
 \frac{a}{H_J^3}=O(\epsilon_{\rm d}^2),\qquad \epsilon_{\rm d}\ll1.
 \label{eq:ordering}
\end{equation}
Two conditions are needed for this ordering. The force must balance the leading velocity term, so $\mathcal G/(H_J^4f_{RR})=O(\epsilon_{\rm d})$. In pure $R^2$ gravity every constant $H_J$ solves the vacuum constraint. In the models considered here, the Einstein--Hilbert term and the variation of the quadratic coupling supply the force. In addition, the dimensionless ratios $R_0f_{RRR}/f_{RR}$ and $R_0^2f_{RRRR}/f_{RR}$ must remain bounded at the orders retained. A sharp feature in $f(R)$ can invalidate the expansion even when $|\dot H_J|/H_J^2$ is small. In that case the Taylor remainder must be checked over the interval between $R_0$ and $R$.

Using $U_R=Rf_{RR}/2$ and $U_{RR}=(f_{RR}+Rf_{RRR})/2$, the second-order equation is
\begin{equation}
 \mathcal E_2=m a+\gamma v-\beta v^2+\mathcal G=0,
 \label{eq:E2}
\end{equation}
with all curvature derivatives evaluated at $R_0$:
\begin{equation}
 m=18H_Jf_{RR},\qquad\gamma=54H_J^2f_{RR},\qquad
 \beta=9f_{RR}-378H_J^2f_{RRR},\qquad
 \mathcal G=3H_J^2f_R-U.
 \label{eq:dictionary}
\end{equation}
The linear coefficient collects three pieces,
\begin{equation}
 \gamma=72H_J^2f_{RR}+18H_J^2f_{RR}-6U_R=(72+18-36)H_J^2f_{RR}=54H_J^2f_{RR},
 \label{eq:gamma_derivation}
\end{equation}
from the explicit velocity term, from $3H_J^2f_R$ and from $-U$. Likewise the $f_{RRR}v^2$ coefficient is $54+432-108=378$, and the potential term supplies $-9f_{RR}v^2$. Keeping only $U_{RR}$ would give the wrong quadratic coefficient. At the next order
\begin{equation}
 \mathcal E_3=\mathcal E_2+\mathcal M\,va+\mathcal K\,v^3,
 \qquad
 \mathcal M=108H_Jf_{RRR},\qquad
 \mathcal K=1188H_J^2f_{RRRR}-36f_{RRR}.
 \label{eq:E3}
\end{equation}
Both new terms are of third order in eq.~\eqref{eq:ordering}; the next inertial term, $324H_Jf_{RRRR}v^2a$, is of fourth order. Solving for the acceleration,
\begin{equation}
 a_2=\frac{-\gamma v+\beta v^2-\mathcal G}{m}.
 \label{eq:a2}
\end{equation}
At third order two forms are equally consistent. One divides by $m+\mathcal Mv$,
\begin{equation}
 a_3^{\rm res}=\frac{-\gamma v+\beta v^2-\mathcal G-\mathcal K v^3}{m+\mathcal M v},
 \label{eq:a3res}
\end{equation}
and so resums part of the fourth-order terms. The strict truncation is
\begin{equation}
 a_3^{\rm str}=a_2-\frac{\mathcal M v a_2+\mathcal K v^3}{m}.
 \label{eq:a3str}
\end{equation}
On the slow branch the two accelerations agree through third order. Over the inflationary intervals tested below, $\mathcal Mv\geq0$ and $m>0$, so the resummed denominator stays positive. For other actions its sign requires a separate check. Section~\ref{sec:numerical} reports both the residual obtained by inserting the exact trajectory into the truncated equation and the error accumulated by evolving that equation.

Within eq.~\eqref{eq:E2}, $\gamma/m=3H_J$, and a local coordinate that removes the explicit quadratic velocity term satisfies
\begin{equation}
 \frac{y_{,H_JH_J}}{y_{,H_J}}=-\frac{\beta}{m},\qquad
 y_{,H_J}\propto H_J^{-1/2}
 [f_{RR}(12H_J^2)]^{7/8},
 \label{eq:general_coordinate}
\end{equation}
on $H_J>0$, $f_{RR}>0$. For $f_{RR}=2\alpha$, eq.~\eqref{eq:E2} is exact and eq.~\eqref{eq:general_coordinate} reduces to the square root of section~\ref{sec:signed_root}.

\subsection{The slow-roll branch and its predictions}
\label{sec:branch}

The truncated equations can be solved order by order on the slow branch, where $\dot H_J$ is a function of $H_J$. Write $v(H_J)=\dot H_J$, so that $\ddot H_J=vv'$ with $v'=\dd v/\dd H_J$. Equations~\eqref{eq:E2} and~\eqref{eq:E3} become first-order equations for $v$, which we solve by iteration:
\begin{align}
 v_1&=-\frac{\mathcal G}{\gamma},\qquad
 v_2=-\frac{\mathcal G-\beta v_1^2+m v_1v_1'}{\gamma},\notag\\
 v_3&=-\frac{\mathcal G-\beta v_2^2+m v_2v_2'+\mathcal M v_2^2v_2'+\mathcal K v_2^3}{\gamma}.
 \label{eq:branch_iteration}
\end{align}
Under the ordering in eq.~\eqref{eq:ordering}, iteration gives successive corrections on the slowly varying branch. The first iterate gives
\begin{equation}
 \epsilon_J^{(1)}=\frac{\mathcal G}{54H_J^4f_{RR}}
 =\frac{3H_J^2f_R-U}{54H_J^4f_{RR}}\bigg|_{R=12H_J^2}.
 \label{eq:general_drift}
\end{equation}
For $f_{RR}=2\alpha$ this is eq.~\eqref{eq:drift}, and $v_2$ and $v_3$ reproduce the two corrections in eq.~\eqref{eq:drift_series}.

From $v(H_J)$ all higher time derivatives follow, for example $\dddot H_J=v\,(vv')'$. The exact frame relations then give the Einstein-frame Hubble-flow functions. With $\epsilon_3\equiv\dot F/(2H_JF)$~\cite{HwangNoh1996,DeFelice2010} and $\dd N_E/\dd t_J=H_J(1+\epsilon_3)$,
\begin{equation}
 \epsilon_{1E}=\frac{3\epsilon_3^2}{(1+\epsilon_3)^2},\qquad
 \epsilon_{2E}=\frac{2\dot\epsilon_3}{H_J\epsilon_3(1+\epsilon_3)^2},\qquad
 \epsilon_{3E}=\frac{\dot\epsilon_{2E}}{H_J\epsilon_{2E}(1+\epsilon_3)},
 \label{eq:flow_E}
\end{equation}
where $\epsilon_{n+1,E}=\dd\ln|\epsilon_{nE}|/\dd N_E$ and $\dot F=f_{RR}(R)\dot R$ uses the full curvature $R=12H_J^2+6v$. The standard second-order expressions~\cite{StewartLyth1993,GongStewart2001,Schwarz2001,Leach2002} then give
\begin{align}
 n_s-1&=-2\epsilon_1-\epsilon_2-2\epsilon_1^2-(2C_{\rm SL}+3)\epsilon_1\epsilon_2-C_{\rm SL}\epsilon_2\epsilon_3,
 &r&=16\epsilon_1(1+C_{\rm SL}\epsilon_2),\notag\\
 n_t&=-2\epsilon_1-2\epsilon_1^2-2(C_{\rm SL}+1)\epsilon_1\epsilon_2,
 &\alpha_s&=-2\epsilon_1\epsilon_2-\epsilon_2\epsilon_3,
 \label{eq:hubble_flow}
\end{align}
with $\epsilon_n\equiv\epsilon_{nE}$ at horizon crossing of the pivot mode and $C_{\rm SL}=\gamma_{\rm E}+\ln2-2\simeq-0.7296$. In the high-curvature quadratic model $\epsilon_3\simeq-\epsilon_J$, and the leading term gives $r\simeq48\epsilon_J^2$, as in eq.~(7.65) of ref.~\cite{DeFelice2010}.

To test this prescription we first locate the pivot on the exact Einstein-frame background. We identify the first crossing of $\epsilon_E=1$, count back 60 e-folds, and choose $k_*=a_{E,*}H_{E,*}$. The frame transformation fixes $H_{J,*}$ at that event. We use this same value in each branch estimate. Taylor arithmetic evaluates derivatives with respect to $H_J$ without finite differences. The coefficients in eq.~\eqref{eq:branch_iteration} use $f$ at $R_0=12H_{J,*}^2$, whereas $\dot F$ in eq.~\eqref{eq:flow_E} uses the reconstructed curvature $R_0+6v_3$. For the six integer values $\delta=0,\ldots,5$ at $(\ell_0,s)=(1,0.01)$, the largest absolute difference from the mode result is $1.03\times10^{-5}$ in $n_s$ and the largest relative difference in $r$ is $0.00510\%$. The branch velocity at the pivot has a relative error below $1.5\times10^{-7}$ for these six couplings. Replacing its flow parameters with the exact ones changes $n_s$ by at most $1.77\times10^{-8}$ and $r$ by at most $2.74\times10^{-7}$ in relative terms. The branch therefore makes a small, measurable contribution, and the second-order spectral formulas account for nearly all of the difference from the mode results. Table~\ref{tab:approx} gives three representative couplings and two power-law tests. Section~\ref{sec:branch_test} separately checks the e-fold count along the branch.

\section{A regularized inverse-logarithmic model}
\label{sec:model}

\subsection{Action, derivatives and potential}

We consider
\begin{equation}
 f(R)=R+\alpha R^2\left[1+\frac{\delta}{\mathcal L(R)}\right],
 \qquad
 \mathcal L(R)=\ell_0+\ln\left(1+\frac{R^2}{R_c^2}\right),
 \label{eq:log_action}
\end{equation}
with $\alpha,R_c,\ell_0>0$ and $\delta\geq0$. At $\delta=0$ this reduces to the Starobinsky action. The logarithm is regular at $R=0$ and $\mathcal L\geq\ell_0$, so the inverse-logarithmic term remains finite as the curvature changes sign. The coefficient of $R^2$ is $\alpha(1+\delta/\ell_0)$ at the origin and approaches $\alpha$ at large $|R|$. The high-curvature form is motivated by a running quadratic coupling~\cite{LiuQuintinAfshordi2026}. The continuation through the origin is a choice of model.

Introduce
\begin{equation}
 x=\alpha R,\qquad s=\alpha R_c,\qquad
 q=\frac{x^2}{x^2+s^2},\qquad
 \mathcal L=\ell_0+\ln(1+x^2/s^2),
 \label{eq:x_s_q}
\end{equation}
and
\begin{equation}
 A=1+\frac{\delta}{\mathcal L}-\frac{2\delta q}{\mathcal L^2},\qquad
 B=1+\frac{\delta}{\mathcal L}-\frac{\delta q}{\mathcal L^2},\qquad
 C=1+\frac{\delta}{\mathcal L}
 -\frac{\delta(5q-2q^2)}{\mathcal L^2}
 +\frac{4\delta q^2}{\mathcal L^3}.
 \label{eq:ABC}
\end{equation}
Direct differentiation gives
\begin{equation}
 F=1+2xB,\qquad f_{RR}=2\alpha C,\qquad
 f_{RRR}=2\alpha^2C_x,\qquad f_{RRRR}=2\alpha^3C_{xx}.
 \label{eq:model_derivatives}
\end{equation}
We also use
\begin{equation}
 \Sigma=\alpha(RF-f)=x^2A,\qquad
 \Theta=\alpha(2f-RF)=x+\frac{2\delta x^2q}{\mathcal L^2},
 \qquad \Sigma_x=2xC.
 \label{eq:S_T}
\end{equation}
The potential and the scalaron mass at the Minkowski minimum are
\begin{equation}
 V_E=\frac{\Mpl^2}{2\alpha}\frac{\Sigma}{F^2},\qquad
 M_{\phi,0}^2=\frac{1}{6\alpha(1+\delta/\ell_0)}.
 \label{eq:log_potential_mass}
\end{equation}

\subsection{Relation to running quadratic gravity}
\label{sec:relation_RG}

The large-number-of-fields approximation of ref.~\cite{LiuQuintinAfshordi2026} has the curvature dependence
\begin{equation}
 \mathfrak f_{\rm RG}(R)\propto
 R^2\left[\lambda_{\rm tH}+\frac{4}{\ln(R^2/\mu_0^4)}\right],
 \label{eq:RG_shape}
\end{equation}
where $\lambda_{\rm tH}$ labels the renormalization-group trajectory and $\mu_0$ is a mass scale. Up to normalization, the quadratic part of eq.~\eqref{eq:log_action} has the same large-curvature form for
\begin{equation}
 \delta=\frac{4}{\lambda_{\rm tH}},\qquad
 \mu_0^2=R_c e^{-\ell_0/2},\qquad
 \mathcal L=\ln\frac{R^2}{\mu_0^4}
 +O\!\left(\frac{R_c^2}{R^2}\right).
 \label{eq:RG_matching}
\end{equation}
The running used in ref.~\cite{LiuQuintinAfshordi2026} follows the proposal of Buccio et al.~\cite{Buccio2024}. Salvio, Strumia and Vitti~\cite{Salvio2026} found that the infrared logarithms interpreted as running couplings depend on the gauge and field parametrization. We use the large-curvature form in eq.~\eqref{eq:RG_shape} as motivation for our choice of $f(R)$. Here the Einstein--Hilbert term is present at all curvatures and the action has a Minkowski minimum. The running couplings, matter loops and Weyl-squared sector of ref.~\cite{LiuQuintinAfshordi2026} are not part of this model, so their strong-coupling bound on $r$ does not carry over.

At inflationary curvatures $|x|\gg s$, the leading logarithm contains $\ell_0$ and $s$ in the combination
\begin{equation}
 \alpha\mu_0^2=s\,e^{-\ell_0/2}.
 \label{eq:mu0}
\end{equation}
At fixed $N_*$ the large-curvature spectra therefore depend primarily on $\delta$ and $\alpha\mu_0^2$. For example, $(\ell_0,s)=(1,0.005)$ and $(1+2\ln2,0.01)$ both give $\alpha\mu_0^2=0.003033$. At $N_*=60$ their scalar tilts differ by less than $2\times10^{-8}$ and their tensor ratios by less than $9\times10^{-7}$ in relative terms. Our reference choice, $(1,0.01)$, gives $\alpha\mu_0^2=0.006065$. Matching through reheating can distinguish continuations with the same $\mu_0$, since the mass in eq.~\eqref{eq:log_potential_mass} depends separately on $\ell_0$.

At fixed $\delta>0$ and $\phi/\Mpl\to\infty$ the potential approaches an inverse-field plateau,
\begin{equation}
 V_E=\frac{\Mpl^2}{8\alpha}
 \left[1-b\frac{\Mpl}{\phi}
 +O\!\left(\frac{\Mpl^2}{\phi^2}\right)\right],
 \qquad b=\sqrt{\frac38}\,\delta,
 \label{eq:inverse_field_plateau}
\end{equation}
because $\mathcal L=\sqrt{8/3}\,\phi/\Mpl+O(1)$ and the leading correction to $V_E$ is $-\delta/\mathcal L$. Slow roll on this plateau gives
\begin{equation}
 N_*\simeq\frac{(\phi/\Mpl)^3}{3b},\qquad
 n_s\simeq1-\frac{4}{3N_*},\qquad
 r\simeq\frac{4\delta^{2/3}}{3N_*^{4/3}}.
 \label{eq:shared_asymptotics}
\end{equation}
With eq.~\eqref{eq:RG_matching} these are the leading results of ref.~\cite{LiuQuintinAfshordi2026}. The limit is not uniform at $\delta=0$, where the exponential Starobinsky plateau takes over.

\begin{figure}[!htbp]
\centering
\includegraphics[width=\linewidth]{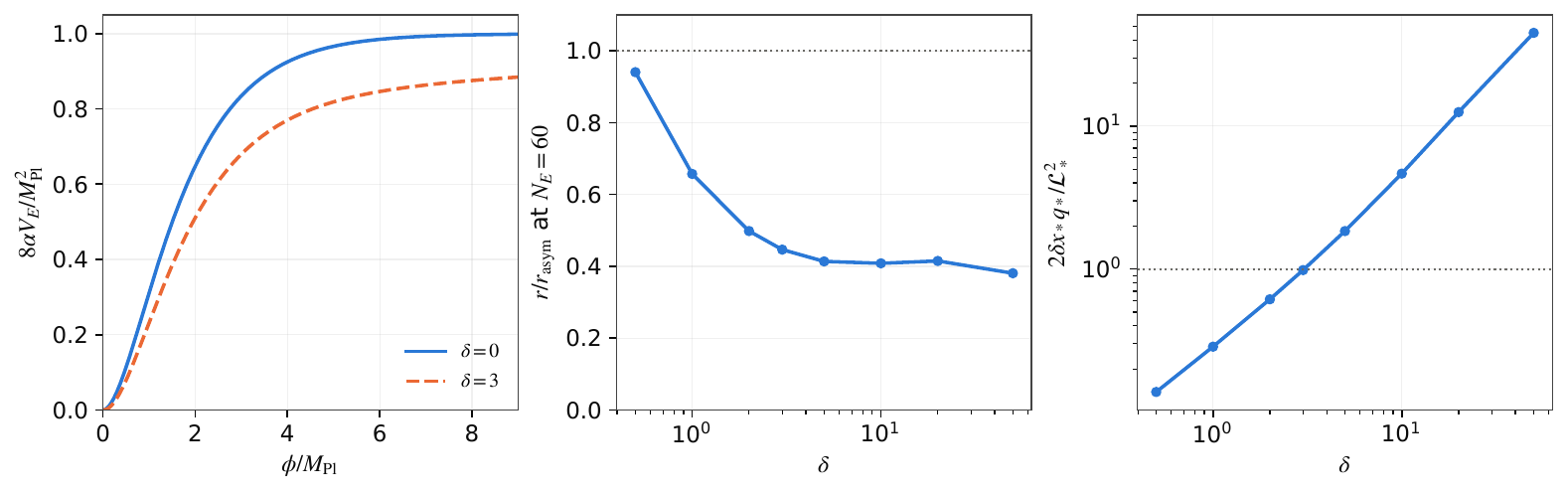}
\caption{Left: Einstein-frame potentials for $\delta=0$ and $\delta=3$ with the same $\alpha$ ($\ell_0=1$, $s=0.01$). Middle: tensor-to-scalar ratio from the mode calculation at $N_E=60$, divided by the asymptotic value of eq.~\eqref{eq:shared_asymptotics}; the ratio would be one if the asymptotic formula were exact. Right: ratio of the logarithmic to the Einstein--Hilbert contribution to the potential slope at the pivot.}
\label{fig:potential_eh}
\end{figure}

At $N_*=60$ the leading large-field formula gives $n_s=0.9778$ and $r=0.0118$ for $\delta=3$. The numerical spectra give $0.9738$ and $0.00527$ (section~\ref{sec:perturbations}). Figure~\ref{fig:potential_eh} helps explain the difference. Although the Einstein--Hilbert term is only $1.6\%$ of the quadratic part of $f$ at this pivot, the two terms contribute almost equally to the slope numerator $2f-RF$. The ratio of the logarithmic contribution to the Einstein--Hilbert contribution is $2\delta x_*q_*/\mathcal L_*^2=0.98$. At $\delta=20$ this ratio rises to 12.5, yet $r/r_{\rm asym}$ remains close to 0.4. Finite-logarithm corrections matter there as well. Over the larger couplings sampled, $\mathcal L_*\simeq18$--21, whereas the leading field-dependent term $\sqrt{8/3}\,\phi_*/\Mpl$ is only about 9--14. The ratio $\delta/\mathcal L_*$ rises from 0.16 at $\delta=3$ to 2.4 at $\delta=50$. At the sampled couplings between 2 and 50 the leading formula overestimates $r$ by a factor of about 2.0--2.6. As $\delta$ tends to zero, $r$ approaches the finite Starobinsky value while $r_{\rm asym}$ vanishes.

\subsection{\texorpdfstring{Positivity of $f_{RR}$}{Positivity of f\_RR}}
\label{sec:positivity}

\begin{proposition}
For $\ell_0=1$, $s>0$ and $\delta\geq0$, eq.~\eqref{eq:log_action} satisfies $f_{RR}\geq2\alpha$ at every real $R$. The function $F(x)$ is strictly increasing and has a unique negative zero $x_b$. On $x>x_b$ the scalaron potential is non-negative and has a Minkowski minimum at $x=0$.
\end{proposition}

\begin{proof}
Write $C=1+\delta P(\mathcal L,q)/\mathcal L^3$ with
\begin{equation}
 P(\mathcal L,q)=\mathcal L^2-(5q-2q^2)\mathcal L+4q^2.
 \label{eq:positivity_polynomial}
\end{equation}
For $1/2<q<1$ the discriminant of $P$, as a quadratic in $\mathcal L$, is $q^2(1-2q)(9-2q)<0$, so $P>0$. For $0\leq q\leq1/2$, $\mathcal L=1-\ln(1-q)\geq1+q$; on this interval $P_{,\mathcal L}\geq2-3q+2q^2>0$ and $P(1+q,q)=(1-q)(1-2q)+2q^3>0$. Thus $C\geq1$. Since $F_x=2C\geq2$, $F$ is strictly increasing; with $F(0)=1$ and $F\to-\infty$ as $x\to-\infty$ it has one zero $x_b<0$. Finally $\Sigma(0)=0$ and $\Sigma_x=2xC$ give $\Sigma>0$ for $x\neq0$, and eq.~\eqref{eq:log_potential_mass} gives the statement about the potential.
\end{proof}

For other values of $\ell_0$, we use $\mathcal L=\ell_0-\ln(1-q)$ to minimize $P/\mathcal L^3$ over $0\leq q<1$. This removes $s$ from the calculation. The numerical threshold above which $P$ remains non-negative is $\ell_P\simeq0.59823$. Below it, positivity of $C$ requires $\delta< -1/\min_q(P/\mathcal L^3)$. The resulting upper limits on $\delta$ are approximately 16.6, 6.26 and 2.95 at $\ell_0=0.5$, 0.4 and 0.3, respectively. Conversely, for $\delta=3$ and 10 the lower thresholds on $\ell_0$ are approximately 0.3021 and 0.4534.
We use $\ell_0=1$ in the reheating calculations. Varying $s$ at this value changes the effective inflationary scale $\mu_0$ while keeping the low-curvature mass fixed at given $\alpha$ and $\delta$.

The potential diverges as $F\to0^+$ because $\Sigma(x_b)>0$. An expanding Einstein-frame solution with finite initial energy cannot reach this boundary, since $\dd\rho_\phi/\dd t_E=-3H_E(\dd\phi/\dd t_E)^2\leq0$.

\subsection{Coefficients and potential slow roll}

At $x_0=12\alpha H_J^2$ eq.~\eqref{eq:dictionary} becomes
\begin{equation}
 m=36\alpha H_J C,\quad \gamma=108\alpha H_J^2 C,\quad
 \beta=18\alpha\left(C-\frac72x_0 C_x\right),\quad
 \mathcal G=3H_J^2+\frac{72\alpha\delta H_J^4q}{\mathcal L^2},
 \label{eq:log_coefficients}
\end{equation}
with $C$, $C_x$, $q$ and $\mathcal L$ at $x_0$. The leading drift~\eqref{eq:general_drift} is
\begin{equation}
 \dot H_J\simeq-\frac{1}{36\alpha C(x_0)}
 -\frac{2\delta H_J^2q(x_0)}{3C(x_0)\mathcal L(x_0)^2}.
 \label{eq:log_drift}
\end{equation}
Potential slow roll gives
\begin{equation}
 \epsilon_V=\frac13\left(\frac{\Theta}{\Sigma}\right)^2,\qquad
 \eta_V=\frac23\left[\frac{F}{2C}\frac{\dd}{\dd x}\left(\frac{\Theta}{\Sigma}\right)
 +\left(\frac{\Theta}{\Sigma}\right)^2\right],
 \qquad
 N_E^{(V)}=3\int_{x_{\rm end,V}}^{x_*^{(V)}}\frac{\Sigma C}{\Theta F}\,\dd x,
 \label{eq:potential_SR}
\end{equation}
with $n_s^{(V)}=1-6\epsilon_V+2\eta_V$, $r^{(V)}=16\epsilon_V$ and $\epsilon_V(x_{\rm end,V})=1$. For $\delta=3$, $\ell_0=1$, $s=0.01$ and $N_E^{(V)}=60$ it gives $x_*^{(V)}=55.9502$ and $x_{\rm end,V}=0.473361$.

At $N_*=60$, $\delta=3$ raises $n_s$ from $0.968$ in the quadratic model to $0.974$. We use this coupling to illustrate how the prediction changes when reheating fixes the pivot.

\section{Background evolution and tests of the truncated equations}
\label{sec:numerical}

\subsection{Numerical method}

We use $\tau_J=t_J/\sqrt\alpha$, $\tau_E=t_E/\sqrt\alpha$, $h_{J,E}=\sqrt\alpha H_{J,E}$ and $p=\dd x/\dd\tau_J$. The exact Jordan system is
\begin{equation}
 \frac{\dd h_J}{\dd\tau_J}=\frac x6-2h_J^2,\qquad
 \frac{\dd x}{\dd\tau_J}=p,\qquad
 \frac{\dd p}{\dd\tau_J}=-3h_Jp-\frac{C_x}{C}p^2-\frac{\Theta}{6C},\qquad
 3h_J^2F=\frac{\Sigma}{2}-6h_JCp,
 \label{eq:Jordan_system}
\end{equation}
where the last relation is imposed initially and monitored. In the Einstein frame, with $u=\dd(\phi/\Mpl)/\dd\tau_E$,
\begin{equation}
 \frac{\dd x}{\dd\tau_E}=\frac{uF}{\sqrt6 C},\qquad
 \frac{\dd u}{\dd\tau_E}=-3h_Eu-\frac{\Theta}{\sqrt6 F^2},\qquad
 h_E^2=\frac13\left(\frac{u^2}{2}+\frac{\Sigma}{2F^2}\right),
 \label{eq:Einstein_system}
\end{equation}
and the two frames are related by
\begin{equation}
 h_J=\sqrt F\left(h_E-\frac u{\sqrt6}\right),\qquad
 p=\frac{uF^{3/2}}{\sqrt6 C},\qquad
 \dd\tau_E=\sqrt F\,\dd\tau_J.
 \label{eq:matching_data}
\end{equation}
The equations hold for any $f(R)$ once $F$, $C$, $\Sigma$ and $\Theta$ are given. We start at $x_i=100$ with the slow-roll drift and the full Friedmann constraint,
\begin{equation}
 W=\frac{\Sigma}{2F^2},\qquad W_{,\varphi}=\frac{\Theta}{\sqrt6 F^2},\qquad
 h_{E,i}^2=\frac{W_i+\sqrt{W_i^2+2W_{,\varphi,i}^2/3}}{6},\qquad
 u_i=-\frac{W_{,\varphi,i}}{3h_{E,i}},
 \label{eq:initial_data}
\end{equation}
which gives $u_i=-0.0035141761$ for $\delta=3$. The reference event lies 60 Einstein-frame e-folds before the first $\epsilon_E=1$ crossing, and the Jordan integration starts from the transformed state there.

We integrate both frames with SciPy's DOP853 solver~\cite{SciPy2020}, using relative and absolute tolerances of $2\times10^{-13}$ and $2\times10^{-15}$ and a maximum step of 0.0125 in each frame's dimensionless time. At 40\,001 common events spanning $0\leq\tau_J\leq160$, the maximum absolute differences in $x$, $h_J$ and $p$ are approximately $3\times10^{-13}$, $8\times10^{-15}$ and $5\times10^{-14}$, respectively. The Jordan constraint residual stays below $10^{-11}$. Increasing both tolerances by a factor of 100 and doubling the maximum step changes the pivot curvature by less than $4\times10^{-13}$.

\subsection{Benchmark background}

\begin{figure}[!htbp]
\centering
\includegraphics[width=\linewidth]{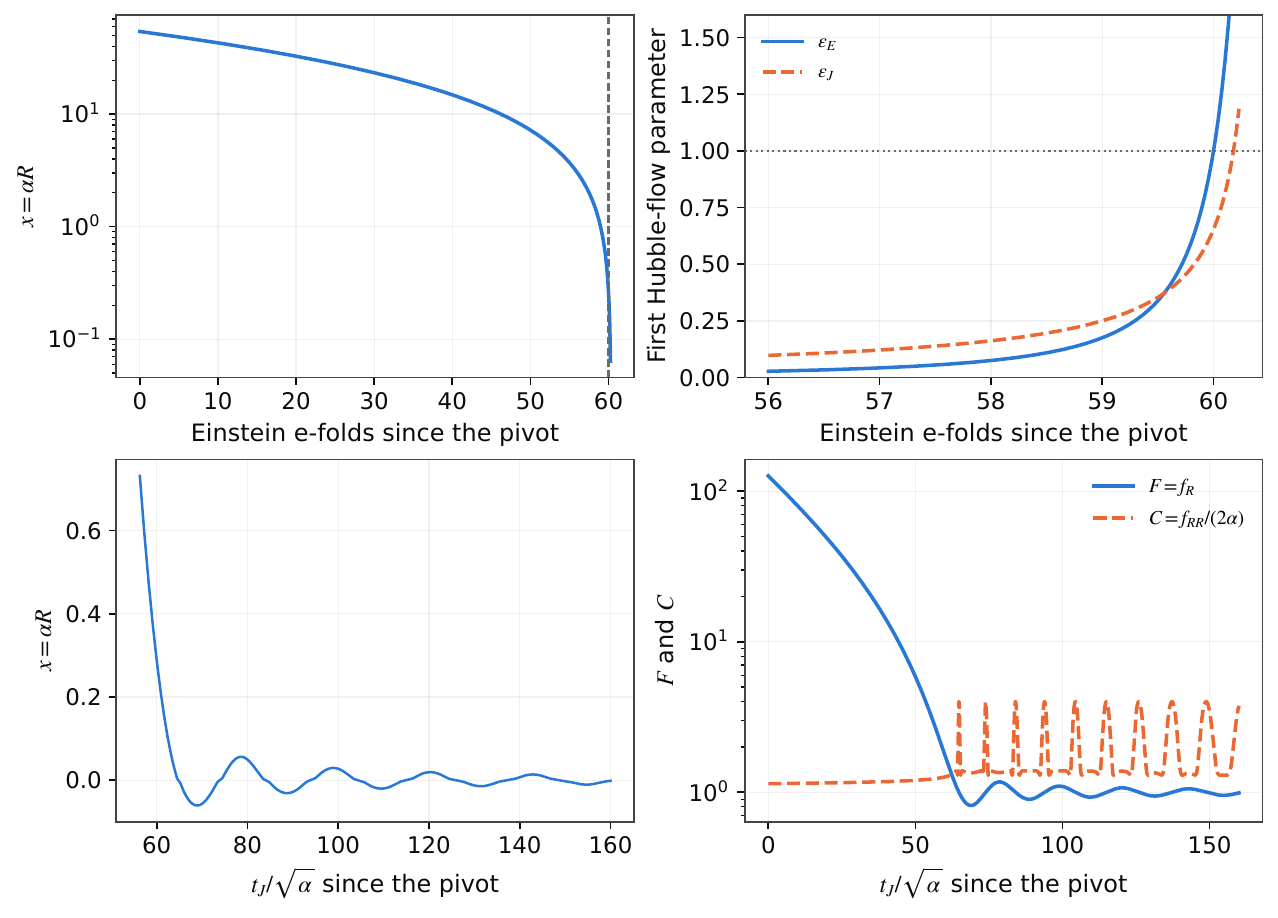}
\caption{Exact vacuum background for $\delta=3$, $\ell_0=1$ and $s=0.01$. Upper panels: curvature during inflation and the first Hubble-flow parameter in each frame near the exit; the e-fold coordinate starts at the reference event. Lower panels: curvature oscillations after inflation and the functions $F$ and $C$, against Jordan time measured from the reference event.}
\label{fig:log_background}
\end{figure}

For $\delta=3$, $\ell_0=1$ and $s=0.01$ the reference state is
\begin{equation}
 x_*=54.2476675,\qquad \phi_*/\Mpl=5.9270815,
 \qquad \epsilon_{E,*}=3.35782\times10^{-4}.
 \label{eq:exact_pivot}
\end{equation}
The first Einstein-frame exit occurs at $x=0.2629670$ and $\tau_J=60.2800$, and the Jordan-frame exit $\epsilon_J=1$ follows $0.1791$ e-fold later. Between any two common events
\begin{equation}
 N_E=N_J+\frac12\ln\frac{F_{\rm end}}{F_*}.
 \label{eq:N_frames}
\end{equation}
After inflation the solution oscillates about $R=0$ (figure~\ref{fig:log_background}). The sampled minima are $F=0.8165$ and $C=1.1397$, in line with section~\ref{sec:positivity}. The oscillations are followed with the exact equations.

\subsection{Truncation tests in the logarithmic model}

\begin{figure}[!htbp]
\centering
\includegraphics[width=\linewidth]{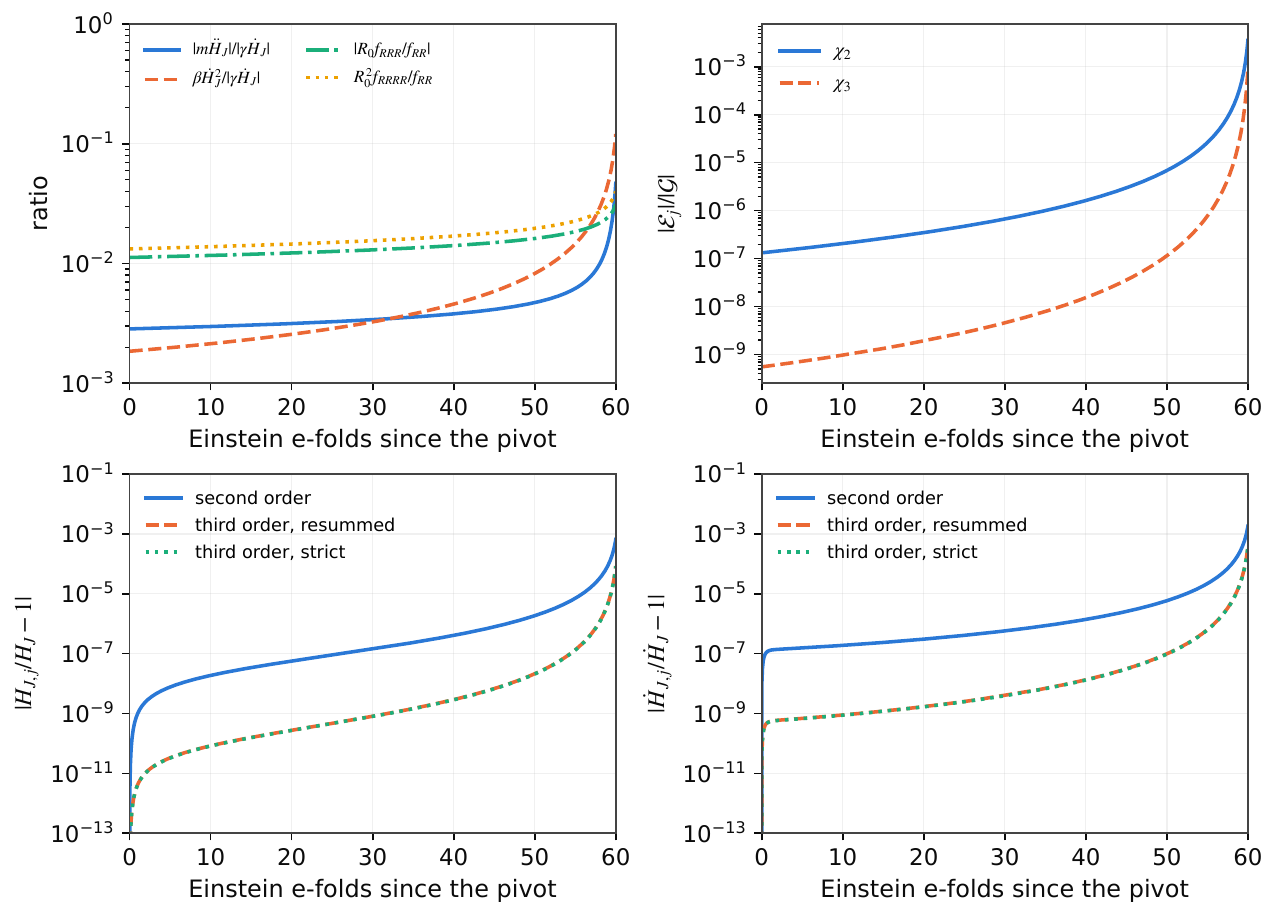}
\caption{Truncation tests for the logarithmic model. Upper left: inertial and quadratic term ratios and the curvature-derivative ratios along the exact background. Upper right: residuals of the second- and third-order equations divided by $|\mathcal G|$. Lower panels: relative errors in $H_J$ and $\dot H_J$ of solutions evolved with eqs.~\eqref{eq:a2}, \eqref{eq:a3res} and~\eqref{eq:a3str}, compared with the exact solution at equal Jordan times.}
\label{fig:truncation}
\end{figure}

At the reference event the second-order ratios and residual are
\begin{equation}
 \frac{|m\ddot H_J|}{|\gamma\dot H_J|}=2.8403\times10^{-3},\qquad
 \frac{|\beta\dot H_J^2|}{|\gamma\dot H_J|}=1.8477\times10^{-3},\qquad
 \chi_2\equiv\frac{|\mathcal E_2|}{|\mathcal G|}=1.318\times10^{-7}.
 \label{eq:log_ratios}
\end{equation}
The omitted mixed term alone is $1.70\times10^{-7}$ of $|\mathcal G|$, and the cubic term also contributes. The curvature-derivative ratios stay small along the whole inflationary trajectory: $R_0f_{RRR}/f_{RR}$ goes from $-0.0112$ at the pivot to $-0.0313$ at the exit, and $R_0^2f_{RRRR}/f_{RR}$ from $0.0132$ to $0.0400$ (figure~\ref{fig:truncation}). Because the logarithm varies slowly, the terms that separate $\mathcal E_3$ from $\mathcal E_2$ are small by construction in this model, and section~\ref{sec:power} adds a model where they are not.

For a pointwise check define
\begin{equation}
 \chi_{j}(t_J)=\frac{|\mathcal E_{j}[H_{J,\rm exact}](t_J)|}
 {|\mathcal G(H_{J,\rm exact}(t_J))|},\qquad j=2,3,
 \label{eq:truncation_tolerance}
\end{equation}
which measures the local residual along the exact solution. For the benchmark, $\chi_2$ first exceeds $10^{-3}$ at $n\simeq59.60$. At the last sample before the Einstein-frame exit we find $\chi_2=3.54\times10^{-3}$ and $\chi_3=8.7\times10^{-4}$.

To test the trajectory itself, we start each truncated equation with the exact pivot values of $H_J$ and $\dot H_J$. We compare at equal Jordan times, using the Einstein-frame e-fold coordinate $n$ of the exact trajectory to label the samples. Table~\ref{tab:truncation_errors} gives the pointwise errors. At $n\simeq30$, the strict third-order error is about 180 times smaller than the second-order error. At $n\simeq60$, the improvement is a factor of eight.

\begin{table}[!htbp]
\centering\small
\caption{Pointwise relative errors $|H_{J,j}/H_J-1|$ of the truncated equations in the logarithmic model ($\delta=3$, $\ell_0=1$, $s=0.01$), at common Jordan times. The labels $n$ are the nearest points of the 40\,001-point sample and differ from the integers by less than $10^{-3}$.}
\begin{tabular}{cccc}
\toprule
$n$ & second order & third order, resummed & third order, strict\\
\midrule
30 & $1.44\times10^{-7}$ & $8.08\times10^{-10}$ & $7.95\times10^{-10}$\\
50 & $1.85\times10^{-6}$ & $2.13\times10^{-8}$ & $2.08\times10^{-8}$\\
55 & $7.07\times10^{-6}$ & $1.34\times10^{-7}$ & $1.31\times10^{-7}$\\
59 & $8.46\times10^{-5}$ & $4.50\times10^{-6}$ & $4.37\times10^{-6}$\\
60 & $6.52\times10^{-4}$ & $8.50\times10^{-5}$ & $8.24\times10^{-5}$\\
\bottomrule
\end{tabular}
\label{tab:truncation_errors}
\end{table}

\subsection{\texorpdfstring{A harder test: $R+\alpha^{p-1}R^p$}{A harder test: R + alpha\textasciicircum (p-1) R\textasciicircum p}}
\label{sec:power}

To see where the expansion fails we use
\begin{equation}
 \alpha f(R)=x+x^p,\qquad 1<p<2,
 \label{eq:power_model}
\end{equation}
a deformation of the Starobinsky model studied in ref.~\cite{Motohashi2015}. For every $R_0>0$, the two curvature-derivative ratios are exactly $p-2$ and $(p-2)(p-3)$, respectively. At large field the potential grows and $\epsilon_V$ tends to $(2-p)^2/[3(p-1)^2]$. Inflation is then close to power law until the Einstein--Hilbert term becomes important. For $p=1.9$, 1.8 and 1.7, we start at $x_i=10^5$, $10^9$ and $10^{15}$, respectively, to allow the background to relax before the pivot. These runs test the derivative expansion and are restricted to $x>0$. We stop at the first Einstein-frame exit. The largest relative difference in $x$ between the two frame integrations is below $5\times10^{-13}$ in these three runs.

Table~\ref{tab:power} shows how the accuracy changes as $\epsilon_{J*}$ increases from 0.040 to 0.179. At $n=30$ the largest second-order error rises from $1.3\times10^{-5}$ to $8.9\times10^{-4}$. The strict third-order errors at the same point are $6.5\times10^{-8}$ and $1.1\times10^{-5}$, respectively. Both third-order forms still improve the trajectory at the exit, although the gain is smaller there. For $p=1.7$ the largest relative error through the exit is $4.4\times10^{-3}$ at second order and $8.7\times10^{-4}$ for the strict third-order equation. As $\epsilon_J$ approaches unity, the derivative counting alone no longer guarantees this accuracy.

\begin{table}[!htbp]
\centering\small
\caption{Truncation tests for $\alpha f=x+x^p$, with the logarithmic model ($\delta=3$) for comparison. Columns give the pivot values of $\epsilon_J$, $R_0f_{RRR}/f_{RR}$ and the residuals $\chi_2$, $\chi_3$, and the largest relative error of $H_J$ on $0\leq n'\leq n$ for the second-order, resummed third-order and strict third-order equations.}
\begin{tabular}{lcccc}
\toprule
Model & $\epsilon_{J*}$ & $R_0 f_{RRR}/f_{RR}$ & $\chi_{2*}$ & $\chi_{3*}$\\
\midrule
log, $\delta=3$ & 0.011 & $-0.011$ & $1.3\times10^{-7}$ & $5.5\times10^{-10}$\\
$p=1.9$ & 0.040 & $-0.100$ & $1.8\times10^{-5}$ & $1.3\times10^{-7}$\\
$p=1.8$ & 0.096 & $-0.200$ & $1.8\times10^{-4}$ & $3.6\times10^{-6}$\\
$p=1.7$ & 0.179 & $-0.300$ & $8.4\times10^{-4}$ & $3.2\times10^{-5}$\\
\bottomrule
\end{tabular}
\par\medskip
\begin{tabular}{llccc}
\toprule
Model & Equation & $n=30$ & $n=59$ & $n=60$\\
\midrule
log, $\delta=3$ & Second order & $1.4\times10^{-7}$ & $8.5\times10^{-5}$ & $6.5\times10^{-4}$\\
 & Third order, resummed & $8.1\times10^{-10}$ & $4.5\times10^{-6}$ & $8.5\times10^{-5}$\\
 & Third order, strict & $7.9\times10^{-10}$ & $4.4\times10^{-6}$ & $8.2\times10^{-5}$\\
\addlinespace
$p=1.9$ & Second order & $1.3\times10^{-5}$ & $2.8\times10^{-4}$ & $2.0\times10^{-3}$\\
 & Third order, resummed & $9.2\times10^{-8}$ & $2.0\times10^{-5}$ & $2.6\times10^{-4}$\\
 & Third order, strict & $6.5\times10^{-8}$ & $1.7\times10^{-5}$ & $2.3\times10^{-4}$\\
\addlinespace
$p=1.8$ & Second order & $1.8\times10^{-4}$ & $3.2\times10^{-4}$ & $3.6\times10^{-3}$\\
 & Third order, resummed & $3.5\times10^{-6}$ & $5.3\times10^{-5}$ & $6.5\times10^{-4}$\\
 & Third order, strict & $1.8\times10^{-6}$ & $4.4\times10^{-5}$ & $5.2\times10^{-4}$\\
\addlinespace
$p=1.7$ & Second order & $8.9\times10^{-4}$ & $9.0\times10^{-4}$ & $4.4\times10^{-3}$\\
 & Third order, resummed & $3.4\times10^{-5}$ & $8.4\times10^{-5}$ & $1.2\times10^{-3}$\\
 & Third order, strict & $1.1\times10^{-5}$ & $8.7\times10^{-5}$ & $8.7\times10^{-4}$\\
\bottomrule
\end{tabular}
\label{tab:power}
\end{table}

\begin{figure}[!htbp]
\centering
\includegraphics[width=\linewidth]{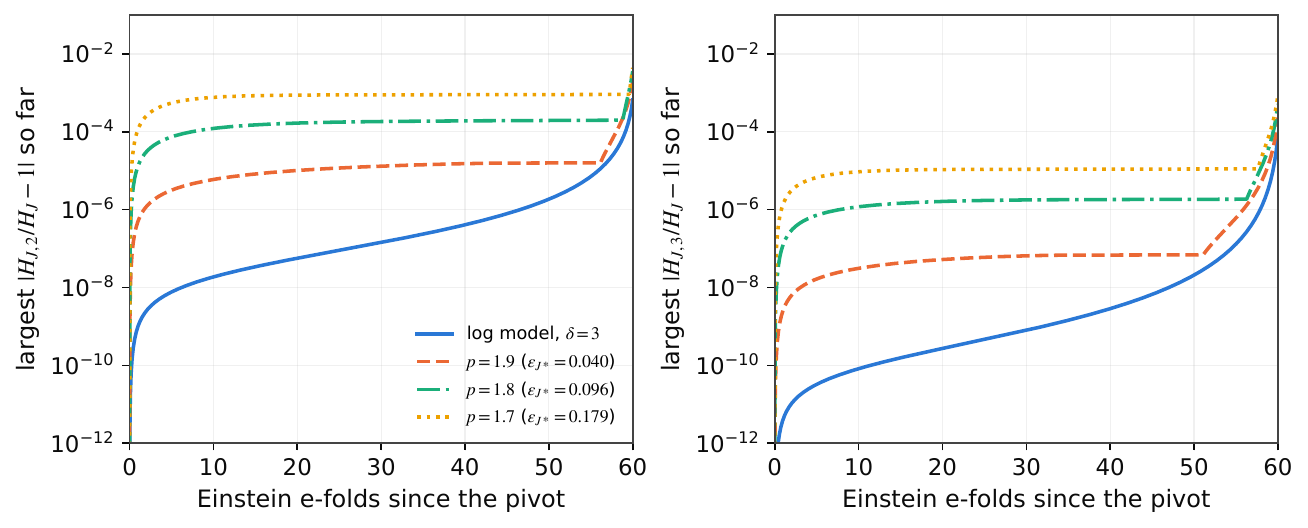}
\caption{Largest relative error of $H_J$ accumulated up to $n$ for the second-order equation (left) and the strict third-order equation (right), for the logarithmic model and for $\alpha f=x+x^p$ with $p=1.9$, $1.8$ and $1.7$. The pivot value of $\epsilon_J$ is given in the legend.}
\label{fig:power}
\end{figure}

\subsection{Accuracy of the slow-roll branch}
\label{sec:branch_test}

At the reference event we compare the branch velocities $v_1$, $v_2$ and $v_3$ of eq.~\eqref{eq:branch_iteration} with the exact $\dot H_J$. For the logarithmic model with $\delta=3$ the relative errors are $10^{-3}$, $6\times10^{-6}$ and $3\times10^{-8}$, so each step gains a factor of about 200. For $p=1.9$ they are $2\times10^{-2}$, $6\times10^{-4}$ and $2\times10^{-5}$, and for $p=1.8$ they are $4\times10^{-2}$, $3\times10^{-3}$ and $2\times10^{-4}$; the gain per step falls as $\epsilon_J$ grows. The branch also counts e-folds. Integrating $\dd N_E=H_J(1+\epsilon_3)\,\dd H_J/v$ along it from the pivot to the exact exit gives $59.83$, $60.04$ and $59.98$ e-folds at the three levels for $\delta=3$, compared with the exact 60. For $p=1.9$ the numbers are $60.88$, $60.06$ and $59.99$.

\section{Perturbations and inflationary predictions}
\label{sec:perturbations}

\subsection{Perturbation equations}

For transverse-traceless tensor perturbations the linear curvature perturbation vanishes, but the evolving $F=f_R$ changes the effective Planck mass. Without anisotropic stress~\cite{Hwang1990,HwangNoh1996}
\begin{equation}
 \ddot h_{ij}+\left(3H_J+\frac{\dot F}{F}\right)\dot h_{ij}
 -\frac{\nabla^2h_{ij}}{a_J^2}=0.
 \label{eq:tensor_eom}
\end{equation}
The propagation speed is one and the extra damping is $\dot F/F$~\cite{Saltas2014,Lombriser2016}. In the quadratic model
\begin{equation}
 \frac{\dot F}{F}=\frac{12\alpha(\ddot H_J+4H_J\dot H_J)}
 {1+12\alpha(\dot H_J+2H_J^2)},
 \label{eq:F_damping}
\end{equation}
and $\dot F/(H_JF)\simeq-2\epsilon_J$ during slow roll at high curvature. The dimensionless damping coefficient is 2.9841 at the pivot and about 2.12 near the Jordan-frame exit (figure~\ref{fig:star_ratios}). Its slow-roll estimate, $3-2\epsilon_J$, loses accuracy as the exit approaches.

The scalar perturbation is described by the comoving curvature perturbation $\mathcal R$, on the rolling branch where comoving scalaron slicing is regular. In the Einstein frame the scalar and tensor canonical modes obey~\cite{Mukhanov1981,Sasaki1986,Mukhanov1988}
\begin{equation}
 v_{s,k}''+\left(k^2-\frac{z_s''}{z_s}\right)v_{s,k}=0,\qquad
 v_{T,k}''+\left(k^2-\frac{a_E''}{a_E}\right)v_{T,k}=0,
 \label{eq:canonical_modes}
\end{equation}
where $z_s=a_E\,\dd\phi/\dd\ln a_E$, primes denote derivatives with respect to conformal time and $\dd\eta=\dd t_J/a_J=\dd t_E/a_E$.

\subsection{Spectra at a fixed number of e-folds}
\label{sec:spectra}

We integrate the modes on the exact Einstein-frame background, with $n=\ln(a_E/a_{E,*})$ as the time variable and $n=0$ placed $N_*$ e-folds before the first $\epsilon_E=1$ crossing. The Bunch--Davies state~\cite{BunchDavies1978,GrivellLiddle1996} is imposed at a finite ratio $\Lambda_{\rm i}=k/(a_EH_E)=400$, and the frozen modes are evaluated at $n=10$. With $\kappa=k\sqrt\alpha$,
\begin{equation}
 \mathcal P_{\mathcal R}=\frac{\kappa^2|\widehat{\mathcal R}|^2}{4\pi^2\alpha\Mpl^2},\qquad
 \mathcal P_T=\frac{2\kappa^2|\widehat h|^2}{\pi^2\alpha\Mpl^2},
 \label{eq:mode_powers}
\end{equation}
where $\widehat{\mathcal R}$ and $\widehat h$ are the amplitudes defined in appendix~\ref{app:convergence}. The tensor power includes both helicities and tends to $2H_E^2/(\pi^2\Mpl^2)$ in the slow-roll limit. The tilts and the running come from centred five-point derivatives in $\ln k$ with a default spacing of $\Delta\ln k=0.02$. We halve the spacing to $0.01$ in the convergence test. Appendix~\ref{app:convergence} shows that varying the numerical settings changes $n_s$ by about $10^{-7}$ or less in the two convergence tests.

Table~\ref{tab:spectral_scan} gives the results at $N_*=60$. The tensor tilt follows the second-order consistency relation: $n_t/(-r/8)$ equals $1+\epsilon_1+\epsilon_2$ to within $1.1\times10^{-3}$. The amplitude fixes the overall scale. If $\mathcal P_{\mathcal R}^{(1)}$ is eq.~\eqref{eq:mode_powers} with $\alpha\Mpl^2=1$, then
\begin{equation}
 \alpha\Mpl^2=\frac{\mathcal P_{\mathcal R}^{(1)}(k_*)}{A_s},
 \label{eq:normalization}
\end{equation}
and we take $A_s=2.1\times10^{-9}$, from $\ln(10^{10}A_s)=3.044$~\cite{Planck2018}.

\begin{table}[!htbp]
\centering\small
\caption{Spectra from the mode calculation at $N_*=60$ for $\ell_0=1$, $s=0.01$. The running $\alpha_s$ and the tensor tilt $n_t$ come from the same five-point stencil as $n_s$.}
\begin{tabular}{ccccccc}
\toprule
$\delta$ & $n_s$ & $r$ & $10^4\alpha_s$ & $10^4n_t$ & $n_t/(-r/8)$ & $10^{-8}\alpha\Mpl^2$\\
\midrule
0 & 0.96775 & 0.003007 & $-5.24$ & $-3.88$ & 1.0335 & $13.06$\\
1 & 0.97027 & 0.003730 & $-5.15$ & $-4.81$ & 1.0308 & $9.997$\\
2 & 0.97226 & 0.004489 & $-5.05$ & $-5.77$ & 1.0286 & $7.917$\\
3 & 0.97384 & 0.005271 & $-4.93$ & $-6.77$ & 1.0269 & $6.444$\\
4 & 0.97511 & 0.006066 & $-4.82$ & $-7.78$ & 1.0255 & $5.365$\\
5 & 0.97613 & 0.006865 & $-4.71$ & $-8.79$ & 1.0244 & $4.552$\\
\bottomrule
\end{tabular}
\label{tab:spectral_scan}
\end{table}

\subsection{Comparison with slow-roll estimates at the same event}
\label{sec:sr_compare}

Table~\ref{tab:approx} compares four estimates at the pivot defined in section~\ref{sec:branch}. The first uses potential slow roll. The next two use the exact background flow parameters in the first- and second-order spectral formulas. The last uses the third-order branch to supply those parameters. For $\delta=3$, first-order potential slow roll underestimates $n_s$ by $6.3\times10^{-4}$ and overestimates $r$ by $2.8\%$. Second-order Hubble flow on the exact background reduces these differences to about $8\times10^{-6}$ and $0.005\%$, respectively.
For the logarithmic model the HF2 and Branch columns agree to the precision shown. At $\delta=3$ the branch velocity has a relative error of about $3\times10^{-8}$. For these benchmarks the truncation error of eq.~\eqref{eq:hubble_flow} accounts for nearly all of the difference from the mode results. This test measures the local prediction once $H_{J,*}$ is known. For $p=1.8$, where $\epsilon_{E,*}\simeq0.02$, the branch adds an error of its own. Its total error is $3.9\times10^{-5}$ in $n_s$ and $0.047\%$ in $r$, whereas second-order Hubble flow with the exact parameters gives $1.8\times10^{-5}$ in $n_s$.

\begin{table}[!htbp]
\centering\small
\caption{Slow-roll estimates evaluated at the exact pivot of the mode calculation ($N_*=60$), relative to the mode results: $\Delta n_s=n_s^{\rm est}-n_s^{\rm modes}$ in units of $10^{-4}$ and $\Delta r/r$ in per cent. ``Pot.'' is first-order potential slow roll; HF1 and HF2 are first- and second-order Hubble flow with the exact $\epsilon_{1,2,3}$; ``Branch'' is HF2 with $\epsilon_{1,2,3}$ from the third-order slow-roll branch.}
\resizebox{\linewidth}{!}{%
\begin{tabular}{l cccc cccc}
\toprule
 & \multicolumn{4}{c}{$10^4\,\Delta n_s$} & \multicolumn{4}{c}{$100\,\Delta r/r$}\\
\cmidrule(lr){2-5}\cmidrule(lr){6-9}
model & Pot. & HF1 & HF2 & Branch & Pot. & HF1 & HF2 & Branch\\
\midrule
log, $\delta=0$ & $-7.35$ & $-3.83$ & $-0.10$ & $-0.10$ & $3.517$ & $2.413$ & $0.004$ & $0.004$\\
log, $\delta=3$ & $-6.30$ & $-3.51$ & $-0.08$ & $-0.08$ & $2.806$ & $1.926$ & $0.005$ & $0.005$\\
log, $\delta=5$ & $-5.79$ & $-3.28$ & $-0.07$ & $-0.07$ & $2.529$ & $1.737$ & $0.005$ & $0.005$\\
$p=1.9$ & $-0.84$ & $-0.41$ & $-0.02$ & $-0.01$ & $0.166$ & $0.115$ & $0.002$ & $-0.002$\\
$p=1.8$ & $8.85$ & $8.86$ & $0.18$ & $0.39$ & $0.001$ & $0.001$ & $0.000$ & $-0.047$\\
\bottomrule
\end{tabular}}
\label{tab:approx}
\end{table}

Potential slow roll gives a different comparison if it also sets the pivot. For $\delta=3$, imposing $\epsilon_V=1$ and counting back 60 e-folds with the potential integral gives $n_s^{(V)}=0.97394$ and $r^{(V)}=0.005222$. These values happen to lie closer to the mode results. On the exact trajectory, the potential end point occurs 0.28 e-fold before the end of inflation, and the interval from the potential pivot to the potential end point is 61.16 e-folds. The potential pivot thus lies 1.44 e-folds earlier than the exact pivot. This shift partly cancels the first-order spectral error.

\subsection{Sensitivity to the reference event and to the renormalization scale}

Table~\ref{tab:spectral_sensitivity} varies the pivot and the effective scale separately at $\delta=3$. Increasing $N_*$ from 50 to 55 raises $n_s$ by 0.0032 and lowers $r$ from 0.00699 to 0.00603. At fixed $N_*=60$, increasing $\mu_0$ raises both observables. The two parameter choices with $\alpha\mu_0^2=0.003033$ agree to the precision shown.

\begin{table}[!htbp]
\centering\small
\caption{Sensitivity of the $\delta=3$ spectra to $N_*$ and to $\alpha\mu_0^2=s\,e^{-\ell_0/2}$.}
\begin{tabular}{cccccc}
\toprule
$N_*$ & $s$ & $\ell_0$ & $\alpha\mu_0^2$ & $n_s$ & $r$\\
\midrule
50 & 0.01 & 1 & 0.006065 & 0.96791 & 0.006985\\
55 & 0.01 & 1 & 0.006065 & 0.97115 & 0.006025\\
60 & 0.005 & 1 & 0.003033 & 0.97338 & 0.004962\\
60 & 0.01 & 1 & 0.006065 & 0.97384 & 0.005271\\
60 & 0.02 & 1 & 0.01213 & 0.97433 & 0.005651\\
60 & 0.01 & $1+2\ln2$ & 0.003033 & 0.97338 & 0.004962\\
\bottomrule
\end{tabular}
\label{tab:spectral_sensitivity}
\end{table}

\subsection{Number of e-folds from scalaron decay}
\label{sec:reheating}

We determine the e-fold number of the pivot $k_*=0.05\,{\rm Mpc}^{-1}$ under a specific reheating assumption. The scalaron oscillates about the Minkowski minimum and decays perturbatively into the four real components of a minimally coupled Standard Model Higgs doublet. Neglecting the Higgs mass, the width is~\cite{GorbunovPanin2011,BezrukovGorbunov2012}
\begin{equation}
 \Gamma=\frac{4M_{\phi,0}^3}{192\pi\Mpl^2},
 \label{eq:decay_rate}
\end{equation}
where $M_{\phi,0}$ is the mass in eq.~\eqref{eq:log_potential_mass}. If the Higgs has curvature coupling $\xi_H$, the tree-level partial width acquires a factor $(1-6\xi_H)^2$, in the convention where $\xi_H=1/6$ is conformal~\cite{GorbunovTokareva2013}. Near this value the Higgs channel is suppressed and other channels, including anomaly-induced decay into gauge bosons, can determine the total width. Changing that width changes the reheating temperature, the inferred $N_*$ and hence $n_s$ and $r$.

We follow the vacuum oscillations for five Einstein-frame e-folds after the first exit. In each e-fold interval we calculate $\int p_\phi\,\dd t_E/\int\rho_\phi\,\dd t_E$. For $\delta=3$ this ratio is 0.13 in the first interval and 0.15 in the second, when the scalaron samples the nonquadratic part of the potential. Its magnitude falls below $10^{-3}$ in the fifth interval. We then evolve the energy density as pressureless matter until $H_E=\Gamma$ and assume immediate thermalization with $g_*=106.75$. Entropy conservation after thermalization gives~\cite{LiddleLeach2003,Planck2018}
\begin{equation}
 N_*=\ln\frac{H_*}{H_0}-\ln\frac{k_*}{a_0H_0}-N_{\rm reh}
 +\ln\frac{T_0}{T_{\rm reh}}+\frac13\ln\frac{g_{s,0}}{g_{s,\rm reh}},
 \label{eq:matching}
\end{equation}
where $N_{\rm reh}$ counts Einstein-frame e-folds from the exit to $H_E=\Gamma$, $H_0=67.4\,{\rm km\,s^{-1}Mpc^{-1}}$, $T_0=2.7255$\,K and $g_{s,0}=43/11$. Since $H_*$ and $\Gamma$ depend on $\alpha\Mpl^2$, which depends on $N_*$ through eq.~\eqref{eq:normalization}, we iterate until $N_*$ changes by less than $2\times10^{-3}$.

Table~\ref{tab:reheating} gives $N_*=51.49$ for $\delta=0$. For comparison, the value 54.37 quoted at $k=0.002\,{\rm Mpc}^{-1}$ in ref.~\cite{BezrukovGorbunov2012} corresponds to about 51.2 at our pivot after subtracting $\ln(0.05/0.002)$. For $\delta=3$ the minimal Higgs width gives $N_*=51.59$, $T_{\rm reh}=2.9\times10^9$ GeV, $n_s=0.9690$ and $r=0.0067$. Increasing or decreasing the width by a factor of ten shifts $N_*$ by about 0.38 e-fold in either direction. Instantaneous reheating gives $N_*=55.9$ and $n_s=0.9717$. Moving the transition to pressureless evolution from five e-folds after exit to four or six changes the inferred $N_*$ by less than $1.4\times10^{-4}$ at fixed spectral normalization.

At $\alpha\mu_0^2=0.006065$, linear interpolation between the results for $\delta=8$ and $\delta=10$ in table~\ref{tab:reheating} gives $\delta\simeq8.7$ and $r\simeq0.012$ at $n_s=0.9743$. The same tilt can move as other parameters vary. For example, at $\delta=3$, increasing $\alpha\mu_0^2$ to $0.0303$ raises $n_s$ from $0.9690$ to $0.9703$.

\begin{table}[!htbp]
\centering\small
\caption{Self-consistent $N_*$ for perturbative decay into a minimally coupled Higgs doublet and the resulting spectra ($\ell_0=1$). $M_{\phi,0}$ and $T_{\rm reh}$ are in GeV.}
\begin{tabular}{cccccccc}
\toprule
$\delta$ & $\alpha\mu_0^2$ & $M_{\phi,0}$ & $T_{\rm reh}$ & $N_*$ & $n_s$ & $r$ & $10^{-8}\alpha\Mpl^2$\\
\midrule
0 & 0.00607 & $3.19\times10^{13}$ & $5.1\times10^{9}$ & 51.49 & 0.9626 & 0.0040 & $9.72$\\
1 & 0.00607 & $2.55\times10^{13}$ & $3.6\times10^{9}$ & 51.52 & 0.9652 & 0.0049 & $7.61$\\
2 & 0.00607 & $2.32\times10^{13}$ & $3.1\times10^{9}$ & 51.55 & 0.9673 & 0.0058 & $6.14$\\
3 & 0.00607 & $2.21\times10^{13}$ & $2.9\times10^{9}$ & 51.59 & 0.9690 & 0.0067 & $5.07$\\
4 & 0.00607 & $2.15\times10^{13}$ & $2.8\times10^{9}$ & 51.62 & 0.9704 & 0.0076 & $4.28$\\
5 & 0.00607 & $2.12\times10^{13}$ & $2.8\times10^{9}$ & 51.66 & 0.9715 & 0.0085 & $3.66$\\
6 & 0.00607 & $2.11\times10^{13}$ & $2.7\times10^{9}$ & 51.69 & 0.9725 & 0.0094 & $3.18$\\
8 & 0.00607 & $2.10\times10^{13}$ & $2.7\times10^{9}$ & 51.74 & 0.9739 & 0.0111 & $2.49$\\
10 & 0.00607 & $2.11\times10^{13}$ & $2.7\times10^{9}$ & 51.78 & 0.9749 & 0.0128 & $2.02$\\
\midrule
3 & 0.00303 & $2.14\times10^{13}$ & $2.8\times10^{9}$ & 51.57 & 0.9685 & 0.0063 & $5.41$\\
3 & 0.0121 & $2.29\times10^{13}$ & $3.1\times10^{9}$ & 51.61 & 0.9696 & 0.0071 & $4.71$\\
3 & 0.0303 & $2.43\times10^{13}$ & $3.4\times10^{9}$ & 51.61 & 0.9703 & 0.0078 & $4.20$\\
\bottomrule
\end{tabular}
\label{tab:reheating}
\end{table}

\subsection{Comparison with data}
\label{sec:data}

\begin{figure}[!htbp]
\centering
\includegraphics[width=\linewidth]{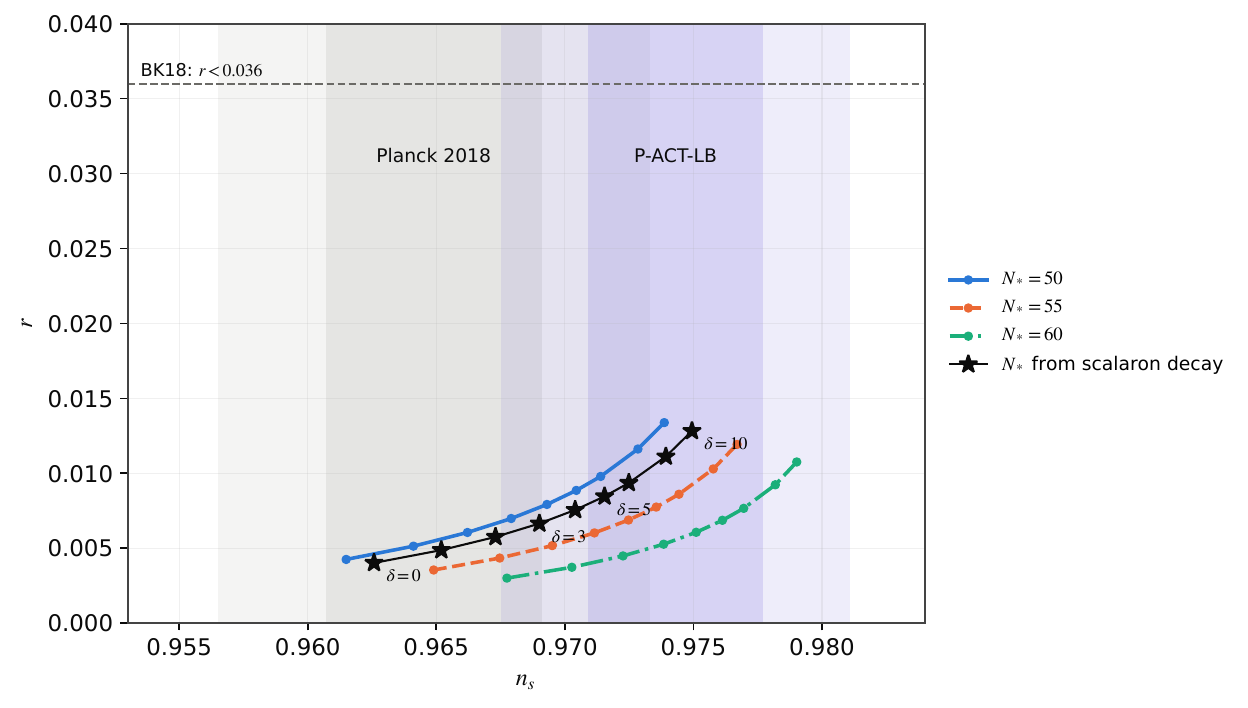}
\caption{The logarithmic model in the $n_s$--$r$ plane for $\ell_0=1$, $s=0.01$ and $\delta=0,1,\dots,6,8,10$ (markers, $\delta$ increasing to the right). Coloured curves: fixed $N_*$. Stars: $N_*$ from gravitational scalaron decay (table~\ref{tab:reheating}). Shaded bands: one and two quoted standard deviations about the marginalized $n_s$ means from Planck 2018 and P-ACT-LB. Dashed line: the BK18 95\% upper limit on $r$.}
\label{fig:nsr}
\end{figure}

Figure~\ref{fig:nsr} compares these predictions with the published one-dimensional constraints. The bands show the central $n_s$ values from Planck 2018 (TT,TE,EE+lowE+lensing)~\cite{Planck2018} and P-ACT-LB~\cite{ACTDR6_2025}, with widths of one and two quoted standard deviations. All plotted tensor ratios lie below the published 95\% limits of $0.036$ (BK18)~\cite{BICEPKeck2021} and $0.038$ (P-ACT-LB-BK18)~\cite{ACTDR6ext_2025}. With minimal Higgs reheating the scalar tilts for $\delta=0$ and 3 lie 3.4 and 1.6 quoted standard deviations below the P-ACT-LB central value. For $3\leq\delta\leq5$ the tilt lies within 1.6 quoted standard deviations of both the Planck 2018 and the P-ACT-LB central values. These distances describe the tilt alone. The intersection at $\delta\simeq8.7$ illustrates the shift with coupling at fixed $\mu_0$ and reheating prescription. Inferring a preferred coupling would require a joint likelihood for the cosmological and model parameters, including assumptions about reheating. Extensions of the cosmological model can also broaden the tilt constraint~\cite{Cheslog2026}.

\section{Conclusions}
\label{sec:conclusion}

The Hubble rate gives a direct description of both slow roll and the subsequent vacuum oscillations in quadratic gravity. When two expanding solutions reach the same decreasing $H_J$, the difference between their velocities contracts at a rate that approaches three per Jordan-frame e-fold during slow roll. After inflation the signed variable $Y$ passes smoothly through zero. Its small-amplitude oscillation gives the scalaron frequency in the curvature and the familiar mean expansion $\langle H_J\rangle\simeq2/[3(t_J-t_c)]$.

For general metric $f(R)$, expanding the Friedmann constraint about $12H_J^2$ gives an approximate velocity at a specified Hubble rate. The distinction between specifying that rate and locating the pivot matters. With the pivot supplied by the exact background, the third-order branch gives $\dot H_J$ at the pivot to $3\times10^{-8}$ for $\delta=3$. Integrating it between the pivot and exit values of $H_J$ taken from the exact background gives 59.98 e-folds, compared with 60 on the exact trajectory. At the six couplings $\delta=0,1,\ldots,5$, its flow parameters, used in the second-order spectral formulas, reproduce the mode results to better than $1.1\times10^{-5}$ in $n_s$ and $0.0052\%$ in $r$. For these six benchmarks, the second-order spectral formulas account for nearly all of the difference from the mode results. The power-law tests show where the expansion loses accuracy. The third-order trajectory remains closer to the exact solution at the exit in all three cases, but the improvement is smaller than it is during slow roll.

The Einstein--Hilbert term has a measurable effect on the slope of the inverse-logarithmic potential even when it contributes little to the action. At $\delta=3$ and $N_*=60$ its contribution to the slope numerator $2f-RF$ is comparable to that of the logarithmic term. The full mode calculation gives $r=0.00527$, compared with $0.0118$ from the leading large-field formula. At larger couplings, finite-logarithm corrections also matter. These results concern the regularized action studied here, whose continuation to low curvature fixes the scalaron mass.

For perturbative decay into a minimally coupled Higgs doublet, the same $\delta=3$ model gives $N_*=51.59$, $n_s=0.9690$ and $r=0.0067$. This tilt is 1.6 quoted standard deviations below the P-ACT-LB value and within one standard deviation of the Planck 2018 value. At fixed $\alpha\mu_0^2=0.006065$, increasing $\delta$ to about 8.7 brings the scalar tilt to the P-ACT-LB central value. A change in the Higgs curvature coupling can shift the pivot, and a change in the renormalization scale can shift the tilt. A measurement of $\delta$ would require those dependences to enter a joint likelihood.

\section*{Acknowledgments}
This work is supported in part by the National Natural Science Foundation of China (NSFC) Grant Nos.\ W2533015, 12375052 and 11935009, and by the Tianjin University Self-Innovation Fund for Extreme Basic Research Grant No.\ 2025XJ21-0007.

\section*{Data and code availability}
\addcontentsline{toc}{section}{Data and code availability}
The supplementary archive supplied with this manuscript contains the Python source, numerical results, figure data and instructions for reproducing the calculations. It includes independent checks of the background equations and perturbation spectra.

\appendix
\section{Perturbation equations and numerical checks}
\label{app:convergence}

\subsection{Quadratic actions}

For $g_{ij}=a_J^2(\delta_{ij}+h_{ij})$ with $h^i{}_i=\partial^ih_{ij}=0$, the quadratic tensor action is~\cite{Hwang1990,HwangNoh1996}
\begin{equation}
 S_T^{(2)}=\frac{\Mpl^2}{8}\int\dd t_J\dd^3x\,a_J^3F
 \left[\dot h_{ij}\dot h_{ij}-\frac{\partial_kh_{ij}\partial_kh_{ij}}{a_J^2}\right],
 \label{eq:tensor_action}
\end{equation}
and on an interval where comoving scalaron slicing is regular the scalar action is
\begin{equation}
 S_{\mathcal R}^{(2)}=\frac12\int\dd t_J\dd^3x\,a_J^3Q_s
 \left[\dot{\mathcal R}^{\,2}-\frac{(\nabla\mathcal R)^2}{a_J^2}\right],
 \qquad Q_s=\frac{3\Mpl^2\dot F^2}{2F[H_J+\dot F/(2F)]^2}.
 \label{eq:scalar_action}
\end{equation}
Both frames have the same conformal time. The tensor pump field is $a_J\sqrt F=a_E$, and the scalar pump satisfies $z_s^2=a_J^2 Q_s=a_E^2(\dd\phi/\dd\ln a_E)^2$. The overall sign of $z_s$ is constant on a rolling branch and drops out of $z_s''/z_s$.

\subsection{Mode equations and initial conditions}

Let $n=\ln(a_E/a_{E,*})$, $\varphi=\phi/\Mpl$, $U_\varphi=\dd\varphi/\dd n$ and $\kappa=k\sqrt\alpha$, with $a_{E,*}=1$ and $\kappa_*=h_{E,*}$. The background equations are
\begin{equation}
 \frac{\dd x}{\dd n}=\frac{U_\varphi F}{\sqrt6 C},\qquad
 \frac{\dd U_\varphi}{\dd n}=-(3-\epsilon_E)
 \left(U_\varphi+\sqrt{\frac23}\frac{\Theta}{\Sigma}\right),\qquad
 \epsilon_E=\frac{U_\varphi^2}{2},\qquad h_E^2=\frac{\Sigma}{2F^2(3-\epsilon_E)}.
 \label{eq:N_background}
\end{equation}
We evolve $\widehat{\mathcal R}=\sqrt{2k}\,\Mpl\mathcal R_k$ and $\widehat h=\sqrt{2k}\,a_E^{-1}v_{T,k}$, which obey
\begin{equation}
 \widehat{\mathcal R}_{,nn}
 +\left(3-\epsilon_E+2\frac{U_{\varphi,n}}{U_\varphi}\right)
 \widehat{\mathcal R}_{,n}+\sigma_k^2\widehat{\mathcal R}=0,\qquad
 \widehat h_{,nn}+(3-\epsilon_E)\widehat h_{,n}+\sigma_k^2\widehat h=0,
 \qquad \sigma_k=\frac{\kappa}{a_Eh_E}.
 \label{eq:N_modes}
\end{equation}
At $\sigma_k=\Lambda_{\rm i}$ the canonical vacuum gives, up to a common phase,
\begin{equation}
 \widehat{\mathcal R}_i=\frac{1}{a_{E,i}U_{\varphi,i}},\qquad
 \widehat{\mathcal R}_{,n,i}=
 \left(-i\Lambda_{\rm i}-1-\frac{U_{\varphi,n,i}}{U_{\varphi,i}}\right)\widehat{\mathcal R}_i,\qquad
 \widehat h_i=\frac{1}{a_{E,i}},\qquad
 \widehat h_{,n,i}=(-i\Lambda_{\rm i}-1)\widehat h_i.
 \label{eq:BD_data}
\end{equation}
This finite-$\Lambda_{\rm i}$ state approximates the short-wavelength vacuum, and we test its dependence on $\Lambda_{\rm i}$ below. The mode solver uses DOP853 with tolerances $2\times10^{-11}$ (relative) and $2\times10^{-12}$ (absolute) and maximum step 0.02 in $n$. The background~\eqref{eq:N_background} uses $3\times10^{-13}$, $3\times10^{-15}$ and step 0.03 and reproduces the cosmic-time reference state.

\subsection{Convergence}

Table~\ref{tab:convergence} changes one numerical setting at a time, starting from the default run used in the main text. Over all rows $n_s$ changes by less than $1.1\times10^{-7}$ and $r$ by less than $3.4\times10^{-6}$ in relative terms. The scalar and tensor Wronskians are conserved to better than $10^{-9}$ from initialization to $n=1$. Evaluating the modes at $n=12$ instead of $n=10$ changes the powers by less than $2.1\times10^{-9}$.

\begin{table}[!htbp]
\centering\small
\caption{Numerical checks of the mode calculation at $N_*=60$. Each row changes one setting of the default run (first row of each block): the initial ratio $\Lambda_{\rm i}$, the wavenumber spacing of the five-point stencil, the solver tolerances, the maximum step, or the evaluation time $n_{\rm eval}$.}
\begin{tabular}{clcc}
\toprule
$\delta$ & setting & $n_s$ & $r$\\
\midrule
0 & default & 0.96774729 & 0.003006547\\
0 & $\Lambda_{\rm i}=100$ & 0.96774735 & 0.003006557\\
0 & $\Lambda_{\rm i}=200$ & 0.96774733 & 0.003006551\\
0 & $\Lambda_{\rm i}=800$ & 0.96774732 & 0.003006547\\
0 & $\Delta\ln k=0.01$ & 0.96774729 & 0.003006547\\
0 & tolerances$/10$ & 0.96774729 & 0.003006547\\
0 & max.\ step$/2$ & 0.96774729 & 0.003006547\\
0 & $n_{\rm eval}=12$ & 0.96774729 & 0.003006547\\
\midrule
3 & default & 0.97384465 & 0.005270992\\
3 & $\Lambda_{\rm i}=100$ & 0.97384475 & 0.005271005\\
3 & $\Lambda_{\rm i}=200$ & 0.97384472 & 0.005270998\\
3 & $\Lambda_{\rm i}=800$ & 0.97384469 & 0.005270993\\
3 & $\Delta\ln k=0.01$ & 0.97384465 & 0.005270992\\
3 & tolerances$/10$ & 0.97384465 & 0.005270992\\
3 & max.\ step$/2$ & 0.97384465 & 0.005270992\\
3 & $n_{\rm eval}=12$ & 0.97384464 & 0.005270992\\
\bottomrule
\end{tabular}
\label{tab:convergence}
\end{table}

\subsection{Independent checks}

We first evolve the canonical variables in eq.~\eqref{eq:canonical_modes} using analytic pump fields. For $\delta=0$ and 3 the relative differences in the scalar and tensor powers from the amplitude calculation are below $8\times10^{-11}$. We then integrate the Jordan-frame background and evolve $\mathcal R$ and the tensor modes with eqs.~\eqref{eq:scalar_action} and~\eqref{eq:tensor_eom}. Both calculations impose the same initial condition at the same event, using $a_EH_E=a_J[H_J+\dot F/(2F)]$. At the two benchmarks, the absolute difference in $n_s$ and the relative difference in $r$ are each below $10^{-10}$. Finally, a power-law background with constant $\epsilon_E=0.01$ provides an analytic test. Its canonical mode is
\begin{equation}
 \sqrt{2k}\,v_k=\sqrt{\frac{\pi y}{2}}\,H_\nu^{(1)}(y),\qquad
 y=-k\eta=\frac{k}{(1-\epsilon_E)a_EH_E},\qquad
 \nu=\frac{3-\epsilon_E}{2(1-\epsilon_E)}.
 \label{eq:powerlaw_validation}
\end{equation}
Starting from this solution at $k/(a_EH_E)=80$, the numerical powers at $n=8$ agree with it to $3.2\times10^{-12}$ and their ratio gives $r=16\epsilon_E$ to rounding accuracy.

The functions $F$, $C$, $C_x$ and $C_{xx}$ of both models were checked against Taylor arithmetic with 75 significant digits at $81$ curvature--parameter pairs; the largest scaled difference is $6.4\times10^{-14}$. An exact rational computation confirms the coefficients of eqs.~\eqref{eq:dictionary} and~\eqref{eq:E3}, and the constraint propagation~\eqref{eq:propagation} holds numerically to $5.2\times10^{-10}$ at random phase-space points of the logarithmic model.

\begin{figure}[!htbp]
\centering
\includegraphics[width=0.62\linewidth]{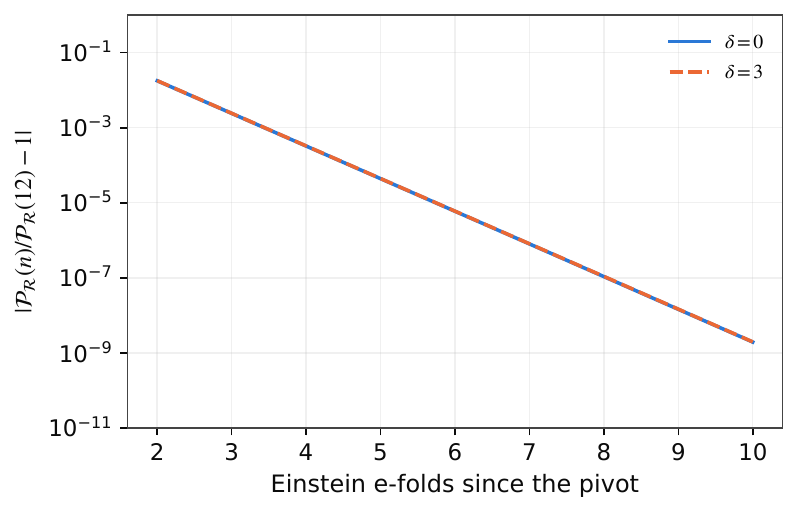}
\caption{Scalar freeze-out for $\delta=0$ and $3$ at $N_*=60$: relative difference between the power at $n$ and at $n=12$ after horizon exit.}
\label{fig:freezeout}
\end{figure}


\end{document}